\documentclass[12pt]{article}
\usepackage{geometry}                
\usepackage{graphicx}
\usepackage{amssymb}
\usepackage{amsmath}
\usepackage{epstopdf}
\usepackage{natbib}
\usepackage{comment}

\usepackage{pgfplots}
\usepackage{bm}
\usepackage{authblk}
\usepackage{hyperref}

\setcitestyle{authoryear,open={(},close={)}}

\newtheorem{thm*}{Theorem}

\newcommand{\beginsupplement}{%
        \setcounter{table}{0}
        \renewcommand{\thetable}{S\arabic{table}}%
        \setcounter{figure}{0}
        \renewcommand{\thefigure}{S\arabic{figure}}%
     }

\title{On Graphical Models and Convex Geometry}
\author[1]{Haim Y. Bar}
\author[2]{Martin T. Wells}
\affil[1]{Department of Statistics, University of Connecticut}
\affil[2]{Department of Statistics and Data Science, Cornell University}
    \date{}                                           

\begin{document}
\maketitle

\begin{abstract}
We introduce a mixture-model of beta distributions to identify significant correlations among
$P$ predictors when $P$ is large. The method relies on theorems in convex geometry, which we
use to show how to control the error rate of edge detection in graphical models. Our `betaMix'
method does not require any assumptions about the network structure, nor does it assume
that the network is sparse.  The results in this article hold for a wide class of data generating distributions that include light-tailed and heavy-tailed spherically symmetric distributions.
\end{abstract}

\section{Introduction}
\subsection{Support discovery and covariance matrix estimation in high-dimensional settings}
Even in the age of `big data' linear regression remains one of the most useful tools available to scientists in all disciplines. The linear regression framework has been built upon the sturdy foundation of the normal
theory, and thus offers powerful inferential and prediction tools. In many applications
the underlying assumptions of regression appear to be reasonable, namely, that the relationship between the predictors and the outcome is linear, and the measurement errors are independently and identically normally distributed
and are uncorrelated with the predictors. However, in order for this theoretical result to be applicable the number of predictors, $P$, cannot exceed the sample size, $n$.
Using conventional notation, where the outcome (response) variable is $y$, and we assume that it is a linear function of $P$ predictors, $x_j$, plus some random (Gaussian) noise, $\epsilon\sim N(0,\sigma^2)$
$$y=\beta_0+\sum_{j=1}^P\beta_j x_j + \epsilon\,.$$
Using matrix notation, the parameter vector is estimated by the ordinary least squares formula:
$$\hat{\bm\beta}=(X'X)^{-1}X'Y\,.$$
If $P>n$, routine estimation of the regression parameters is not possible since the inverse of the matrix $X'X$ does not exist, and we say that $\bm\beta$ is unidentifiable. Even if $n>P$, inference about $\bm\beta$ may be impractical when $P$ is sufficiently large because standard errors are often large and the width of the confidence interval grows with $P$. For example, Hotelling's $T^2$ yields confidence intervals with width which is proportional to $\sqrt{[P(n-1)F_{P,n-P,\alpha}]/[n(n-P)]}\,.$

To deal with the fact that many modern applications involve a large number of putative predictors and often a modest sample size, statisticians had to develop variable selection methods capable of identifying the true predictors, while limiting the number of irrelevant predictors from being included in the regression model. Arguably, most famous among such methods is the LASSO \citep{Tibshirani:1996}. Such methods assume sparsity, and require that $\log(P)=o(1)$. For example, \cite{vandegeer2014} denote the active set of variables by $S_0=\{j:\beta_j\ne 0\}$ and its cardinality by $s_0$. To prove their main result (Theorem 2.2) they further require that the $n$ samples are i.i.d. Gaussian, $X'X$ has a strictly positive smallest eigenvalue, and that $(X'X)^{-1}$ is row-wise sparse: $\max_js_j=|\{k\ne j:(X'X)^{-1}\ne 0\}|=o(n/\log(P))$. Under these assumptions \cite{vandegeer2014} derive asymptotic confidence intervals for the LASSO estimator, $\hat{b}_{LASSO}$ and obtain an $o_{\mathbb{P}}(1)$ estimator for the precision matrix, $\Sigma^{-1}$.

\cite{Wainwright2009} establishes precise conditions on $P$, $s_0$ (adopting the previous notation), and $n$, the sample size needed to recover the sparsity pattern using the LASSO. One consequence of his analysis is that under the assumptions of $s_0$-sparsity of the true $\bm\beta$ and invertibility of the matrix $X_{S_0}'X_{S_0}$, the LASSO estimator converges to $\bm\beta$ in the $\ell_2$ norm if $\log(P)/n=o(1)$. 
In Corollary 2 of his main result \cite{Wainwright2009} shows that for standard Gaussian designs (a) the LASSO can only recover $\bm\beta$ with support cardinality $s_0\le (1+o(1))n/2\log(P)$ where $n=\nu P$ for some $\nu\in(0,1)$, and it fails with probability converging to 1 if there exist  $c_2 > c_1 > 0$ such that $s_0\in(c_1P, c_2P)$; and (b) if $k=\alpha P$ for some $\alpha\in(0,1)$ then the LASSO requires a sample size $n>2\alpha P\log[(1-\alpha)P]$ in order to obtain an exact recovery of $\bm\beta$.

\cite{Reid2016} propose a sparse regression and marginal testing approach for data with correlated predictors. They first cluster the predictors, and then take the most informative predictor in a cluster as the `prototype'. They then apply either the LASSO or marginal significance testing to the much smaller set of predictors which were selected as prototypes. 
\cite{Efron2004} introduce the popular LARS method (least angle regression), which tackles the same problem. The idea is essentially similar to forward stepwise selection. Initially all $\beta_j=0$ and in the first iteration the variable most correlated with the response is selected and its coefficient is set according to the sign of its correlation with the response. In each step, the current estimator is used to update the residuals, and the selected $\beta_j$ is increased in the direction of the sign of its correlation with $y$, until some other predictor $x_{k}$ is as correlated with the residual vector  as $x_j$. The process is repeated -- $(\beta_{j},  \beta_{k})$ are increased in their joint least squares direction, until another predictor $x_k$  is as correlated with the residual, and so forth, until all the predictors are in the model. With a simple modification to the algorithm, \cite{Efron2004} show that the LARS algorithm yields all the LASSO solutions.

In the literature mentioned thus far the main objective was to recover the vector of regression parameters, $\bm\beta$. Many authors have extended the scope to the case in which the covariance matrix in the large-$P$ setting also has to be estimated. This is important in a number of applications, including dimension reduction via principal component analysis (PCA) or singular value decomposition (SVD), spatial analysis, classification via discriminant analysis, and fitting graphical models. \cite{bickel2008} consider a method based on hard-thresholding and show that if $\log P/n\rightarrow 0$ and the true covariance matrix is sparse in the sense that in each row at most $c_0(P)\ll P$ elements are non-zero, then the threshold estimate is consistent. Furthermore, the rate of convergence of their estimator is shown to be $O_{\mathbb{P}}[c_0(P)(\log(P)/n)^{(1-q)/2}]$, for $q\in[0,1)$.
\cite{BickelYan2008} consider the importance of sparsity in covariance matrix estimation. They define sparsity in terms of points lying on or near a low dimensional sub-manifold of a $P$-dimensional space.  They consider sparsity in the covariance matrix or in the precision matrix, so that in either case each row in the matrix is sparse in the operator norm, that is, the number of non-zero elements in each row in the covariance matrix is small (less than some $s$). \cite{BickelYan2008} discuss properties of the estimator for the `true' dimension of the data, which is assumed to be much smaller than $P$.

Sparse estimation of the covariance estimation and covariance selection has been studied extensively in recent years. See, for example,  \cite{levina2008sparse}, \cite{rothman2008sparse}, and \cite{warton2008penalized}. \cite{cai2013two} consider testing equality of covariance matrices and support discovery in two-sample, high-dimensional and sparse settings, and \cite{zhu2017testing} use high-dimensional covariance matrices tests to detect schizophrenia risk genes.
In the context of graphical models, a common approach is to identify edges in a high-dimensional graph by using the LASSO $P$ times, each time taking another variable as the response and performing variable selection on the other $P-1$. See, for example \cite{meinshausen2006high,friedman2008sparse, yuan2007model, peng2009partial, khare2015convex}.  The computational complexity of these penalized approaches is polynomial in $P$. In contrast, correlation screening methods that are mentioned in the next section are non-iterative algorithms so the computational complexity is of the order $P\log P$ \citep{hero2015foundational}.

\subsection{Our Approach -- the Beta Mixture Model}
While applying the linear model with Gaussian errors to high-dimensional problems has been a natural step which yielded extraordinary advances, both in theory and in applications, it also required making strong assumptions, including sparsity of the mean vector, and the covariance (or precision) matrix. In the original, small-$P$ setting, the assumption that the predictors are uncorrelated seemed reasonable, but correlation between columns of $X$ is inevitable when $P$ is large, and for the most part, the approaches to deal with such correlations have relied on a somewhat ad-hoc two-stage approach, where in the first step a dimension reduction is performed (e.g., by clustering) in order to restore at least in part the validity of the requirement that $X'X$ is invertible. Another motivation for extending the linear model framework to the large-$P$ setting has been the interpretability of the results, namely, that `a unit change in some significant predictor, $x_j$ is associated with $\beta_j$ units increase in $y$'.

However, in many cases involving a large number of predictors there is no reason to think that the relationship between $Y$ and $X$ is linear. For example, a quantitative trait may depend on the expression of many genes in an intricate way, so that we cannot use statements like `holding all other variables constant', and we cannot draw conclusions like `an increase in expression of gene $j$ is associated with $\beta_j$ units increase in the trait', because a change in the expression of that gene may not occur without a simultaneous change in many other genes. For the same reason the  $\bm\beta$-sparsity assumption may not be valid, and the covariance matrix may not be sparse, either. It is quite possible, and in fact common, that a trait is associated with hundreds or even thousands of genes. Such is the case if genes form a highly connected network, which may be necessary because the trait requires the production of many different proteins or it may be evolutionary beneficial as a way to protect against mutations. It may also be the case that the assumption of underlying low dimensionality is not valid. For example, if the predictors have an auto-regressive structure, $AR(m)$, as is the case if the predictors represent repeated measurements (for example, daily log-returns of stocks). It is possible to reduce the dimensionality by taking representative predictors, but doing so results in loss of information about the most prominent feature of the data, namely, its $AR(m)$ structure.

Since in high-dimensional setting neither a linear relationship between $X$ and $Y$, nor the uniqueness of $\bm\beta$, nor its sparsity are some laws of Nature, but rather, mathematically convenient assumptions, we propose to change the perspective and consider obtaining the whole network structure as the main objective, where nodes represent variables and edges represent strong associations between pairs of variables. While the graphical models in the literature aim to do just that, they almost always rely on sparse models. As stated earlier, detecting edges in a network via the graphical LASSO involves designating each predictor in its turn as the response and regressing on the other $P-1$ variables. Our approach, which is described in the next section, does not require sparsity assumptions, and achieves the network structure in a single step. We may, however, choose to treat one variable as a `response', and use our method to perform variable selection by identifying all the nodes (variables) connected via an edge to the response node in the graph.

Like LARS, our method uses partial correlations, but in a very different way. LARS is a stepwise process in which each step involves updating the coefficients of the regression and the residuals, and recalculating the correlations between the remaining predictors and the residuals. Our method calculates all the pairwise correlations just once.  LARS is a variable selection method, used when the goal is to find which predictors are associated with a response variable, while our method finds all the connections between all variables simultaneously. Furthermore, LARS relies on $\bm\beta$-sparsity, while our method does not. There is, however, a more subtle difference between our method and LARS. LARS uses an inclusion criterion, adding variables to the model while a cumulative threshold has not been exceeded, and that threshold depends on a tuning parameter, usually obtained via cross validation. In contrast, our method is based on an exclusion criterion which uses the distribution of pairwise correlations under the null hypothesis (which is discussed in the next section). Thus, our method provides an inferential framework, which is used to control the error rate even in the presence of massive multiple testing.

Our method is based on ideas and results from convex geometry, some of which may seem counter intuitive at first glance. The relevant theorems are stated in Section \ref{sec:method}, but for a comprehensive (and very enjoyable) introduction to convex geometry see \cite{Ball97anelementary} and \citealt[Chapter~2]{blum2020foundations}. The latter reference contains applications to modern data science challenges.
The key to our method is `flipping' the roles of variables and observations and treat the data as $P$ points in $\mathbb{R}^n$, so that each predictor is characterized by $n$ samples. The classical approach views data as $n$ points in $\mathbb{R}^P$, and in the high dimensional setting where $P > n$, all the $n$ points lie on a low-dimensional hyperplane in $\mathbb{R}^P$. This degeneracy causes difficulties for classical statistical methods.  However, if we view the data as $P$ vectors in $\mathbb{R}^n$, then such degeneracy problem no longer exists. This is the mathematical structure that underlies the asymptotics in the high-dimensional-low-sample-size framework developed by \cite{hall2005geometric}.

A convex geometry foundation also allows us to establish a remarkable asymptotic relationship between $n$ and $P$. Specifically, whereas in the aforementioned literature it is required that $\log(P)/n=o(1)$, with our approach as long as $\log(P)$ is $O(n)$, we can detect nonnull edges with high accuracy, while controlling the error rate (and no tuning via cross validation is needed).

Ideas from convex geometry have been applied in the statistics literature to developing thresholds for correlation screening.  In correlation screening the objective is to select variables whose
maximal correlation exceeds a given threshold.  \cite{hero2011, hero2015foundational} developed a novel threshold for marginal correlation screening in the high-dimensional-low-sample-size setting using spherical cap calculations. Their threshold is derived as asymptotic expressions for the mean number of correct discoveries. These expressions depend on a Bhattacharyya measure \citep{basseville1989distance} of average pairwise dependency of the $P$ multivariate scores defined on $\mathcal{S}^{n-2}$.   \cite{hero2011} give $(1 - c_n (P-1)^{-\frac{2}{n - 4}})^{\frac{1}{2}}$ (for $c_n$ equal to the volume of $\mathcal{S}^{n-2}$) as useful correlation screening threshold.  A similar threshold, $(1- P^{-\frac{2}{n - 2}})^{\frac{1}{2}}$, was also developed in \cite{zhang2017spherical} for detecting spurious correlations and low rank correlation structure also by using a spherical cap packing perspective.  Note that both of these thresholds are of the same order of $(1 - P^{ -{\frac{2}{n}}} )^{\frac{1}{2}} = (1-\exp\{-{\frac{2}{n}} \log{P}\})^{\frac{1}{2}} \sim (1- (1-{\frac{2}{n}} \log{P}))^{\frac{1}{2}} = ({\frac{2}{n}} \log{P})^{\frac{1}{2}}$ which is connected to the classical rate of convergence mentioned above.

\cite{cai2011limiting, cai2012phase, cai2013distributions} take an approach to screening correlations via the analysis of discovering minimal pairwise angles. These articles give a very careful analysis of the normalizing constants for the convergence to the particular Weibull-type extremal distribution using spherical cap calculations.  They consider the different asymptotic phase transition regimes where $\log{P}/n \rightarrow \{0, \rm{a \; constant}, \infty \}$.   The phase transitions that are developed are similar in spirit to those in \cite{hero2015foundational} and \cite{zhang2017spherical}.

\cite{Reverter2008} also use partial correlations combined with information theory for the reconstruction of gene co-expression networks. Our approach is more similar to that of \cite{10.1371/journal.pone.0246945} who also considered a mixture model for detecting significant correlations, but their approach relies on Fisher's Z-transformed correlations and their asymptotic normal distribution under the null hypothesis. The nonnull edges in their edgefinder method are modeled as two lognormal distributions - one for significantly positive correlations and one for negative ones. The edgefinder method performs very well especially when $n$ is sufficiently large, but the approach presented here does not require a normalizing transformation, and controlling the error rate relies on a general convex geometry theory. 

The idea of flipping the roles of predictors and observations appeared in the context of variable selection in \cite{SEMMS}, although there the motivation was to improve computational efficiency via the Woodbury identity when the true number of predictors is much smaller than $P$. The approach presented here is more general in that it does not depend on sparsity, nor does it require to define one of the variables as the response. Therefore, the same convex geometry principles can be used to explain the excellent performance of SEMMS. It should be noted that as a variable selection method, SEMMS has been extended to the generalized linear models framework, as well as to quantile regression in the $\bm\beta$-sparse, high-dimensional setting, whereas the beta mixture presented here assumes that the variable are drawn from a spherically symmetric distribution. We briefly discuss these, and other possible extensions in Section \ref{sec:disc}.

Using a key distribution result in Theorem \ref{thm1} (Sec. \ref{sec:method})
we know that pairs of uncorrelated predictors will be nearly perpendicular with high probability if $n$ is sufficiently large. The null distribution of the squared sine of angles between random pairs is $Beta((n-1)/2, 1/2)$ and we could use this fact to detect edges in a graph, while controlling the error rate as frequentists (and note that the result from Theorem \ref{thm1} is very strong and applies to \textit{any} pair, so we do not have to correct for multiple testing). However, we propose an empirical Bayes approach, and use a mixture of two beta distributions where the nonnull component is $Beta(a,b)$ and $a, b$ are estimated from the data. This allows us to get more power, and check model adequacy. It also allows us to adapt the model to situations where the samples are not i.i.d., and we do so by replacing the $(n-1)/2$ parameter in the null component by $(\nu-1)/2$, where $\nu\in(1,n)$ is the `effective sample size'. In many situations, the i.i.d. assumption is unrealistic, and using $Beta((n-1)/2, 1/2)$ as the null distribution will lead to many false discoveries.





%

\section{Method}\label{sec:method}
\subsection{Background -- Convex Geometry Results}

Most intuition about geometry which is based on two and three dimensions can often be misleading in high dimensions.  Specifically, the field of statistics leans heavily on various orthogonal decompositions through the notion of Pythagorean right angles (in Greek, {\it ortho gonia}). From classical linear algebra perspective, the minimal number of orthogonal basis vectors needed to specify an object in a Euclidean space defines its orthogonal dimension. Recent work in convex geometry \citep{kainen2020} seeks to extend the notion of dimension to $\epsilon$-quasi-orthogonal dimension of $\mathbb{R}^n$.  The concept of quasi-orthogonal dimension is obtained by relaxing exact orthogonality so that angular distances between unit vectors are constrained to a closed symmetric interval about $\pi /2$.  For $ \epsilon \in [0, 1]$ a subset of $A \subset \mathcal{S}^{n-1}$ is a $\epsilon$-quasi-orthogonal subset if $x \neq y \in A \Rightarrow |\langle x, y \rangle| \leq \epsilon$. The  $\epsilon$-quasi-orthogonal dimension of $\mathbb{R}^n$ is defined as $\dim_{\epsilon} (n) :=\max\{ |X|: X \subset \mathcal{S}^{n-1}, x \neq y \in X \Rightarrow |\langle x, y \rangle| \leq \epsilon\}$.  Equivalently, the maximum number of nonzero vectors whose pairwise angles lie in the interval [$\arccos(\epsilon), \arccos(-\epsilon)$] centered at $\pi /2$ or the maximum cardinality of an $\epsilon$-quasi-orthogonal subset of $\mathbb{R}^n$.  

\cite{kainen2020} showed that an exponential number of such quasi-orthogonal vectors exist as the Euclidean dimension increases, specifically $\dim_{\epsilon} (n) \geq \exp(n \epsilon^2 /2)$.   The argument for the existence of such large quasi-orthogonal sets comes from packing spherical caps into the surface of $\mathcal{S}^{n-1}$. The spherical caps consist of all points on the sphere within a fixed angular distance from some point, that for $y \in \mathcal{S}^{n-1}$ and $\epsilon> 0$, $\mathcal{C}(y,\epsilon)  := \{x \in \mathcal{S}^{n-1},  \langle x, y \rangle \geq \epsilon\}$.  It is known \citep[p.11]{Ball97anelementary} that the Lebesgue measure of $\mathcal{C}(y,\epsilon)$ is bounded above by $\exp(-n \epsilon^2 /2)$.  The proof of the $\dim_{\epsilon} (n)$ bound then follows from a maximum packing argument.  It is of special note that the two bounds are reciprocals.

The $\exp(-n \epsilon^2 /2)$ upper bound on the Lebesgue measure of $\mathcal{C}(y,\epsilon)$ is quite counter-intuitive since for any fixed $\epsilon$, the bound becomes very small as $n$ increases. Hence, in high dimension, most of the area of the sphere lies very close to its equator.  This is an incidence of the concentration of meaure phenomena. \cite{donoho2000high} notes that increases in dimensionality can often be helpful to the asymptotic analysis. The blessings of dimensionality includes the concentration of measure phenomenon where certain random fluctuations are very well controlled in high dimensions.

Despite the widespread and applicability in statistical modeling, linear subspaces suffer from the drawback that they cannot be analyzed using Euclidean geometry. Indeed, subspaces  of $\mathbb{R}^n$ lie on a special type of Riemannian manifolds, the Grassmann manifold, which has a nonlinear structure. The Grassmann manifold $\mathbb{G}_{n, k}$ is used to study the geometry of the space of all $k$ dimensional subspaces of $\mathbb{R}^n$.  $\mathbb{G}_{n, k}$ is isomorphic to the quotient set $\mathbb{O}(n)/(\mathbb{O}(k) \times \mathbb{O}(n - k))$, where $\mathbb{O}(j)$ denotes the orthogonal group of order
$j$  \citep{absil2009optimization}.

The manifold $\mathbb{G}_{n, k}$ has an invariant measure \citep{james1954normal, lv2013impacts} which can be used to calculate the volumes of sets which are specified in terms of the principal angles $\theta_i$ between $k$ and $l$ dimensional subspaces of $\mathbb{R}^n$. The principal angles between subspaces are the generalization of the concept of the angle between lines. Let $\cal{U}$ and $\cal{V}$ be two subspaces in $\mathbb{G}_{n, k}$ and $\mathbb{G}_{n, l}$ ($k \leq l$)  having a set of principal angles $(\theta_1,\ldots, \theta_k)$, with $\pi/2 \geq\theta_1 \geq\cdots\geq\theta_k \geq0$ and corresponding $k$ pairs of orthogonal unit vectors $(u_{i}, v_{i})$. By setting $\rho_i = \cos\theta_i$ gives the canonical correlations $(\rho_1,\ldots, \rho_k)$ and corresponding pairs of canonical variables $\{u_{i}, v_{i}\}_{i=1}^{k}$ \citep{lv2013impacts}.  The chordal distance between $\cal{U}$ and $\cal{V}$ is $ (\sum_{i = 1}^k \sin^2 \theta_i)^{1/2}$ \citep{conway1996packing} and the maximum chordal distance is $\sin \theta_1$ \citep{absil2009optimization}.

The invariant measure of  the principal angles $(\theta_1,\ldots, \theta_k)$ can be constructed by viewing $\mathbb{G}_{n, k}$ as $\mathbb{V}_{n, k}/\mathbb{O}(k)$, where $\mathbb{V}_{n, k}$ denotes the Stiefel manifold of all orthonormal $k$-frames in~$\mathbb{R}^n$ \citep{lv2013impacts}. By deriving the exterior differential forms on those manifolds, \citet{james1954normal} gave an expression for the invariant (uniform) measure of the principal angles $(\theta_1,\ldots, \theta_k)$ for ($k \leq l$) 

\begin{equation}
\label{measure_kl} 
d \mu_{k,l}^n = C_{k, l}^n 
\prod_{i = 1}^k (\cos^2\theta_i)^{(l-k)/2}
\prod_{i = 1}^k (\sin^2\theta_i)^{(n -l-k)/2}
\prod_{1 \leq i < j \leq k} \bigl(\sin^2
\theta_i - \sin^2 \theta_j\bigr) \,d \theta_1 \cdots d \theta_k
\end{equation}
over $\Theta= \{(\theta_1,\ldots, \theta_k):  \pi/2 > \theta_1 >
\cdots> \theta_k > 0\}$.  The normalization constant is given by
\begin{equation}
\label{norm1} C_{k, l}^n = \prod_{i = 1}^{k}
\frac{A_{k - i+1}  A_{k - i+1} A_{n- l - i +1}}{2 A_{n - i +1}},
\end{equation}
where $A_j = 2 \pi^{j/2}/\Gamma(j/2)$ is the area of the unit sphere $S^{j -1}$.

In the special case where $k = l =1$ the Grassmann manifold $\mathbb{G}_{n, 1}$ is a generalization of the projective space $\mathbb{P}^{n-1}$ corresponding to the lines passing through the origin of the Euclidean space \citep[pg.30]{absil2009optimization}.  The chordal distance between two lines is the sine of their angle.  On $\mathbb{G}_{n, 1}$ the invariant measure $\mu_{1, 1}^n$ has a simple expression for the density of the canonical angle $\theta$  \citep{absil2006largest, lv2013impacts}
\begin{equation}\label{cossqtheta}
\nu_{1}^n (\theta) = \frac{1}{B(\frac{1}{2}, \frac{n-1}{2})} (cos^2\theta)^{-1/2} (1-cos^2\theta)^{(n-1)/2-1}, 
\end{equation}
where $B(a, b) = \Gamma (a)\Gamma (b)/\Gamma (a+b)$ is the beta function. This nice result implies that the $cos^2\theta$ has a $Beta(1/2, (n-1)/2)$ distribution or equivalently $sin^2\theta$ has a $Beta((n-1)/2, 1/2)$ distribution. These are precisely the measure the underlie the spherical cap calculations discussed in the previous section \citep{hero2011, hero2015foundational, cai2011limiting, cai2012phase,  cai2013distributions, zhang2017spherical} that was used in the development of correlation screening rules.

A random vector $V \in \mathbb{R}^P$ is spherically symmetric if, for any orthogonal transformation ${\cal O}$, ${\cal O}V$ has the same distribution as $V$. The class of spherical distributions generalize the multivariate normal distribution and includes the Laplace, logistic, symmetric stable, and $t$- distributions as well as the family of scale mixtures of normal distributions. The connection between spherical symmetry and uniform distributions on the unit sphere is through the equivalence: a random vector $V \in \mathbb{R}^p$ has a spherically symmetric distribution  if and only if $V$ has the stochastic representation $V \stackrel{\mathrm{D}}{=} \| V \| \, U$ where $Pr[ \| V \| = 0]=0$, $U$ and $V$ are independent, and $U$ is uniformly distributed on the unit sphere.  Note that $V / \| V \|$ has the same distribution for the entire family of spherically symmetric distributions. If  $\mu \in \mathbb{R}^P$, $\Sigma$ is a positive definite $P \times P$ matrix, and $V$ is spherically symmetric, then  $X \stackrel{\mathrm{D}}{=} \mu + \Sigma^{{1/2}}Z \in \mathbb{R}^P$ has an elliptically symmetric distribution, ${\mathcal E}_P (\mu, \Sigma)$.   See Chapter 4 of \cite{fourdrinier2018shrinkage} for further details on spherical and elliptical distributions.  Throughout this article we assume the data $\{X_i\}_{i=1}^{n}$ follow an elliptically symmetric distribution  ${\mathcal E}_P (\mu, \Sigma)$.

With the geometric perspective one can use the invariant measure $\mu_{k,l}^n$ to calculate the volumes and probabilities of sets containing the principal angles in the Grassmann manifold \citep{lv2013impacts}.  The case when $k=1$ the result in (\ref{measure_kl}) seems to have appeared in many places and forms (e.g., \cite{muirhead1982} and \cite{watson1983statistics}) but we give the version by Theorem 1.1 in \cite{Frankl1990}. 

\begin{thm*} \label{thm1} 
Let $K$ be a fixed 1-space (line) in $\mathbb{R}^n$, and let $L$ be a random
$l$-space  in $\mathbb{R}^n$. Let $\theta$ be the angle between $K$ and $L$. The
random variables $cos^2\theta$ and $sin^2\theta$ have the beta distributions
$Beta(l/2, (n-l)/2)$ and $Beta((n-l)/2, l/2)$, respectively.
\end{thm*}

In the following we let both $H$ and $L$ be lines ($k=l=1$, that is, are elements of $\mathbb{G}_{n, 1}$), in which case Theorem \ref{thm1} or a change of variables in (\ref{cossqtheta}) yields
\begin{equation}\label{sinsqtheta}
Z\stackrel{\mathrm{def}}{=}\sin^2\theta\sim Beta\left(\frac{n-1}{2},\frac{1}{2}\right)\,.
\end{equation}

Using an asymptotic approximation for the beta function and bounding the beta probability density function in (\ref{sinsqtheta}), Theorem 3.1 in \cite{Frankl1990} gives an approximation for the cumulative distribution function of $\theta$ that holds for any $\alpha\in(0, \pi/2)$
\begin{equation}\label{sinsqcdf}
Pr[\theta \leq \alpha] = Pr[Z \leq \sin^2\alpha] = [(\pi(n-1)/2 )^{1/2}\cos\alpha]^{-1}(\sin\alpha)^{n-1} + o(1)
\end{equation}
\noindent where $o(1) \rightarrow 0$ as $n \rightarrow \infty$.

\begin{figure}
\begin{center}
\begin{tikzpicture}[scale=0.8,
    declare function={
            gamma(\z)=2.506628274631*sqrt(1/\z)+ 0.20888568*(1/\z)^(1.5)+ 0.00870357*(1/\z)^(2.5)- (174.2106599*(1/\z)^(3.5))/25920- (715.6423511*(1/\z)^(4.5))/1244160)*exp((-ln(1/\z)-1)*\z;
        },
        declare function={
            beta(\x,\y)=gamma(\x)*gamma(\y)/gamma(\x+\y);
        },
    declare function={
        fdst(\x,\n) = 1/beta( (\n-1)/2, 0.5) * \x^((\n-1)/2-1) * (1-\x)^(-0.5);
    }
]
\begin{axis}[
    axis lines=left,
    enlargelimits=upper,
    samples=100,
    xmin=0, ymin=0, ymax=15,
    domain=0.001:0.999,
    legend style={at={(0.25,1)},anchor=north,legend cell align=left}
]
legend style={at={(0.03,0.5)},anchor=west}
\addplot [very thick,blue] {fdst(x,10)}; \addlegendentry{$n=10$}
\addplot [very thick,red,dashed] {fdst(x,100)}; \addlegendentry{$n=100$}
\addplot [very thick,brown,dotted] {fdst(x,500)}; \addlegendentry{$n=500$}
\end{axis}
\end{tikzpicture}
\caption{The $Beta\left(\frac{n-1}{2},\frac{1}{2}\right)$ distribution for $n=10$ (solid), 100 (dashed),
and 500 (dotted).}
\label{betadist}
\end{center}
\end{figure}
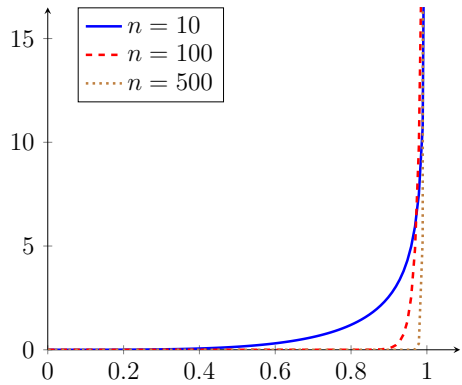

\begin{figure}
\begin{center}
\begin{tikzpicture}[scale=0.8,
    declare function={
        f(\n,\angle) = log10(sqrt(3.14159*(\n-1)/2)*cos(deg(\angle))/sin(deg(\angle))^(\n+1));
    }
]
\begin{axis}[
    axis lines=left,
    enlargelimits=upper,
    samples=100,
    xmin=10, 
    domain=10:250,
    legend style={at={(0.25,1)},anchor=north,legend cell align=left},
    xlabel=$n$, ylabel=$\log_{10}m_\alpha(n)$,  
    ]
legend style={at={(0.03,0.5)},anchor=west}
\addplot [very thick,brown,dotted] {f(x,3.141593/4)}; \addlegendentry{$\alpha=\pi/4$}
\addplot [very thick,red,dashed] {f(x,3.141593/3)}; \addlegendentry{$\alpha=\pi/3$}
\addplot [very thick,blue] {f(x,0.8*3.141593/2)}; \addlegendentry{$\alpha=0.8\pi/2$}
\end{axis}
\end{tikzpicture}
\caption{The logarithm of the number of lines in $\mathbb{R}^{n}$ going through the origin 
such that any two of them determine an angle greater than $\alpha$, for $\pi/4$ (dotted), $\pi/3$ (dashed),
and $0.8\pi/2$ (solid).}
\label{minpalpha}
\end{center}
\end{figure}
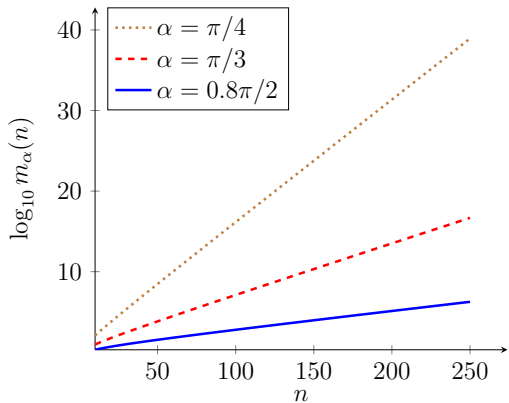

A consequence of (\ref{sinsqcdf}) is that two random vectors are approximately perpendicular with high probability, if the dimension of the space is sufficiently large. If $n$ is large enough, the distribution of $Z$ is highly concentrated close to 1. This phenomena is what is anticipated from the concentration of measure results in \citep[p.11]{Ball97anelementary}.  For example, Figure \ref{betadist} shows the distribution of $Z$ for $n=10$, 100 and 500.  Even with $n=10$ it is very unlikely that $z$ will be less than, say, 0.6,
which means that the probability of two \textit{random} vectors in $\mathbb{R}^{10}$ will be correlated
by chance, is very small.
 In other words, even for moderate values of $n$, $\mathbb{R}^{n}$ is a pretty big space which allows for a lot of random vectors to be sufficiently far from one another (in terms of chordal-based distance). 
Formally, citing Theorem 3.2 from  \cite{Frankl1990} with a slight change in notation, we have the following:
\begin{thm*} \label{thm2}
For any $\alpha\in(0, \frac{\pi}{2})$, there exist more than 
$$m_\alpha(n)=(\pi(n-1)/2)^{1/2}\cos\alpha(\sin\alpha)^{-(n-1)}$$
lines in $\mathbb{R}^{n}$ going through the origin $O$ such that any two of them determine an
angle greater than $\alpha$.
\end{thm*}
The number of lines going through the origin that can be drawn randomly in $\mathbb{R}^{n}$ 
such that the angle between each pair is at least $\alpha$ grows exponentially with $n$. This phenomena is what is foreseen from quasi-orthogonality results in \cite{kainen2020}. Theorem \ref{thm2} 
is illustrated in Figure \ref{minpalpha}, which shows the asymptotically linear relationship between
$\log_{10}m_\alpha(n)$ and $n$ for three
angles: $\alpha=\pi/4, \pi/3$ and $0.8\pi/2$.

We emphasize that Theorem 3.2 in \cite{Frankl1990} refers to \textit{any} pair among
at least $m_\alpha(n)$ randomly chosen lines.
For example, with $n=100$ there are at least 12,713,167 lines that can be drawn randomly
so that the angle between any two lines is at least $60^\circ$.

Now consider the sine of the principal angles between two $k$ dimensional subspaces. \citet{absil2006largest} use the Gauss hypergeomtric function $ {_2F_1}$ with a matrix argument to give the density of the largest principal angle between two random subspaces.  We give the necessary definitions of the Gauss hypergeomtric function in the Appendix.  Using Theorem 1 in  \citet{absil2006largest} and the transformation $z=\sin^2 \theta_1$ we have the multivariate analog of (\ref{sinsqtheta}).

\begin{thm*} \label{thm3} 
Let $K$ and $L$ be two $k$ plane in $\mathbb{R}^n$. Let $\theta_1$ be the largest principal angle between $K$ and $L$. Then the random variables $Z=sin^2\theta_1$ has a probability density function
\begin{equation}
\frac{k(n-k)}{2} \frac{\Gamma(\frac{k+1}{2})\Gamma(\frac{n-k+1}{2})}{\Gamma(\frac{1}{2})\Gamma(\frac{n+1}{2})} z^{k(n-k)/2 -1} (1-z)^{-1/2} {_2F_1}\left(\frac{n-k-1}{2}, \frac{1}{2};\frac{n+1}{2}; z I_{k-1}\right).
\end{equation}
\end{thm*}

\subsection{The Beta-Mixture Method} Consider a situation in which we obtain $ P$ 
quantitative characteristics (predictors) of $n$ random subjects, and assume that $P$ is large, 
possibly much larger than $n$. Such datasets have become very common in
recent years due to advances in high-throughput technologies which allow researchers
to obtain, for example, RNA sequencing data for tens of thousands of genes.

Many graphical model methods regard the data as $n$ points in $\mathbb{R}^P$,
and apply a variable selection method to a multivariate linear (normal) model in order to
detect strong relationship between pairs of predictors.

Our approach is different in that we consider the data as $P$ points in $\mathbb{R}^n$, and
rather than representing each subject by $P$ quantitative characteristics, we view each 
predictor as a point which is determined by a sample of $n$ subjects. `Null' predictors correspond to 
randomly drawn points in $\mathbb{R}^n$. 
Using Theorem 1.1 from \cite{Frankl1990}, pairs of null predictors will be nearly perpendicular
 with high probability, if $n$ is sufficiently large. 
 
 Therefore, our approach to detecting edges in the 
graphical model is to exclude all edges corresponding to pairs of approximately perpendicular
vectors (predictors) in $\mathbb{R}^n$. We use the distributional properties of randomly drawn
vectors to establish statistical properties, and to control the probability of erroneously retaining an
edge in the graph.

The known distribution of null edges allows us to use either a frequentist or a Bayesian inferential
procedure.
Let $\theta_j$ be the angle between the $j$th pair of predictors, $j=1,\ldots,P(P-1)/2$
and let $z_j=\sin^2\theta_j$. Denote the $\epsilon$ quantile of the 
$Beta((n-1)/2,1/2)$ distribution by $Q_\epsilon$.
With a frequentist approach, the $j$th edge in the graphical model exists if the screening rule $z_j<Q_\epsilon$ holds.
For example, suppose that $P=500$ and $n=70$. Since the total number of possible 
edges is 124,750, to control the probability of Type I error (detecting an edge which should not be
in the graph) we may set $\epsilon=10^{-5}$.
In this example, we have $Q_\epsilon(34.5, 0.5)\approx 0.75$, which means that we include an edge in the graph if the angle (modulus 90) between the corresponding pair of vectors is less than $60^\circ$, which is equivalent to a correlation coefficient of at least 0.5 between the two predictors.

The screening rules in \cite{hero2011, hero2015foundational, cai2011limiting, cai2012phase,  cai2013distributions, zhang2017spherical} are based on the maximum correlation exceeding a threshold so these rules are defined by  $\ell_{\infty}$ balls (hypercubes). There are interesting differences between the volume of a hypercube with unit length sides and the volume of a unit radius sphere in high dimensions \citep{blum2020foundations}.  As the dimension of the unit cube increases, its volume is always one and the maximum possible distance between two points grows as the square root of the dimension.  Cast in terms of a confidence set, the $\ell_{\infty}$ ball has fixed volume but a diverging width. In contrast, as the dimension of a unit sphere increases, its volume goes to zero exponentially \citep{Ball97anelementary} and the maximum possible distance between two points stays fixed. \cite{efron2006minimum} pointed out that confidence regions should be constructed to minimize volume. Consequently, in high dimensions confidence regions based on hypercubes, as in screening, may be problematic as they have exponentially larger volume than those based on spherical rules.

An alternative to the screening methodology is to use an empirical Bayes two-group approach.  We define a  mixture model (hereafter called the \textit{betaMix} model)
$$\ell(z_j)= m_{0j}f_0(z_j)+(1-m_{0j})f(z_j)$$
where $m_{0j}$ is a random indicator which is equal 1 if the pair of predictors
corresponding to $z_j$ is approximately perpendicular, and
\begin{eqnarray*}
f_0(z_j)&=&\frac{1}{B(\frac{n-1}{2},\frac{1}{2})}z_j^{\frac{n-1}{2}-1}(1-z_j)^{-\frac{1}{2}}\\
f(z_j)&=&\frac{1}{B(a,b)}z_j^{a-1}(1-z_j)^{b-1}\\
m_{0j}&\sim&Ber(p_0)\,.
\end{eqnarray*}
The mixture  also follows a beta distribution, with parameters
$$m_{0j}\frac{n-1}{2}+(1-m_{0j})a \, \textrm{ and } \, \frac{m_{0j}}{2}+(1-m_{0j})b\,.$$
Note that the alternative distribution, $f(z)$, is very flexible. We do not impose any restrictions on the parameters
$a$ and $b$, except that they have to be positive.
Since the mixture indicator variables are latent, we use the EM algorithm \citep{EM} to estimate the model parameters.
In the M-step we obtain the maximum likelihood estimates for $a$ and $b$. 
In the E-step we replace $m_{0j}$ with their posterior mean,
$$\hat{m}_{0j}=\frac{p_0f_0(z_j)}{p_0f_0(z_j)+(1-p_0)f(z_j)}\,.$$
For $p_0$ we obtain the maximum likelihood estimate, namely, the mean of the Bernoulli random variables, $\hat{p}_0=\bar{m}_{0\cdot}$.
This process is repeated iteratively until convergence is achieved.
We say that an edge in the graph exists if the posterior null probability (under $f_0$)  is smaller than some threshold,
$\hat{m}_{0j}<\tau$ (e.g., $\tau=0.01$).

\subsection{Adaptation to Dependent Samples}\label{depdata}
Parameter estimates for the beta-mixture model in the previous sub-section were derived under the assumption that the $n$ samples are independent, but it may not always be the a reasonable assumption. When this assumption is invalid it is possible, at least conceptually, to incorporate a certain dependence structure into the model and derive a distribution for the null set. One conceivable way to achieve this is to employ a random effects model which accounts for within-cluster correlations, and estimate the intraclass correlation coefficient (ICC). The ICC is often used to determine the \textit{effective} sample size (ESS) of an experiment, and when the ICC is large, the ESS is much smaller than the actual sample size. 

Rather than specifying a possibly incorrect dependence structure we propose a different approach, and instead we model the \textit{consequence} of dependence among observations, namely, a \textit{smaller effective sample size}. Let $\nu\le n$ be the unknown ESS, and let the null distribution be
\begin{eqnarray*}
f_0(z_j)&=&\frac{1}{B(\frac{\nu-1}{2},\frac{1}{2})}z_j^{\frac{\nu-1}{2}-1}(1-z_j)^{-\frac{1}{2}}\,.
\end{eqnarray*}
Note that the second parameter of the null distribution is still 1/2 because the null hypothesis is still that the $P$ vectors are drawn randomly, which means that the results from \cite{Frankl1990} apply, but because observations may be dependent the dimension may be less than the sample size. The ESS $\nu$ is estimated from the data in the M-step of the EM algorithm. The estimating equations for the three parameters, $a$, $b$, and $\nu$ are given in the Appendix.

\subsection{Improving the Convergence of the Estimation Algorithm}
We can improve the convergence of the algorithm without significantly increasing the error rate,
 by carefully changing the support of the non-null component to $[0,C_\delta]$ where 
 $C_\delta\le 1$,  so that the probability
that any $z_j$ for a non-null pair falls in $(C_\delta,1]$ is negligible.  

Let $\delta$ be a small value, and let $M_c=\#\{z_j|z_j> c\}$ be the number of 
pairs for which $z_j$ is greater than some $c$. We choose $C_\delta$ so that
 $(1-\delta)\cdot100\%$ of $M_c$ are from the null distribution:
$$C_\delta=\arg\min_c\left[(1-\delta)M_c - \frac{1}{B(\frac{\nu-1}{2},\frac{1}{2})}\int_c^1z^{(\nu-1)/2-1}(1-z)^{-1/2}dz\right]^2\,.$$

The changes to the iterative estimation algorithm are trivial. We just use a three-parameter beta distribution for the
non-null component: 
$$f(z)=\frac{1}{C_\delta^{a+b-1}B(a,b)}z^{a-1}(C_\delta-z)^{b-1}\,,$$
and update $C_\delta$, in addition to $a$ and $b$ (and $\nu$ if we allow for dependence) in each iteration.

\section{Simulations}
The method described in the previous section has been implemented as an R package called \texttt{betaMix}.
We simulated data with different numbers of predictors, sample sizes, and correlation structures, and in each  configuration we ran the \texttt{betaMix} function and evaluated the goodness of fit of the mixture-model and the ability of the method to recover the true correlation structure in terms of true- and false-positive edges that it detects.
Table \ref{simresults} shows representative results from our simulations. In all cases, data were generated from a multivariate normal distribution with $P$ variables and $N$ samples, and we varied the correlation structure of the normal distribution.
For example, in configurations 1-8 the correlation matrix had a block-diagonal clustered structure, with cluster sizes set to 25, 100, or 500. All the variables within a cluster were correlated, and the correlation coefficient between each pair in a cluster was set to either low (0.3) or high (0.9). Pairs not belonging to the same cluster were set to be uncorrelated.
In scenarios 9-18 we used a band matrix, with band-width set to either 30 or 150. Pairs of variables which correspond to cells within the band were set to be correlated (low or high), whereas outside the band the correlations were set to 0.
The third family of examples in the table (lines 19-24) consists of correlation matrices with a block-diagonal cycle structure, with cycle length being 25 or 50.  For example, in scenario 19 there were 20 blocks along the diagonal, each of dimension 25 by 25, so that the first variable is correlated with the second, the second with the third, and so on, and the 25th variable is correlated with the first variable in the block. 
Six additional correlation structures are described in the Supplementary Material, for a total of 124 simulation configurations. Each configuration was used to generate 30 datasets, and we count the number of correctly and falsely detected edges in each run. An edge is correctly detected if and only if the corresponding cell in the actual correlation matrix was not zero.

The results show that as $N$ increases the probability of detecting a true edge approaches 1. The number of false positive edges is very small, since our model allows to control the error rate. We set the FDR threshold to 0.01, and in all cases the observed error rate was lower. When the magnitude of the correlation coefficient is high, the power of our method increases, and even when $N$ is much smaller than $P$, if $\rho$ is 0.9 our method detects well over 90\% of the edges in all scenarios. Even with a modest correlation coefficient our method has very good power, as a result of the mixture model which allows to borrow strength across pairs of variables.

Note that our model does not assume sparsity, and performs just as well in the non-sparse cases. For instance, in scenarios 5-8 $P=1000$ and there are two clusters, each with 500 variables. This shows that \texttt{betaMix} is very effective in fitting the stochastic block model. In fact, since our model makes no assumptions about the structure of the correlation matrix, nor on the sparsity, it is equally effective when blocks (or cycles, hubs, etc.) overlap. 

\begin{figure}[htbp]
\begin{center}
\includegraphics[scale=0.75]{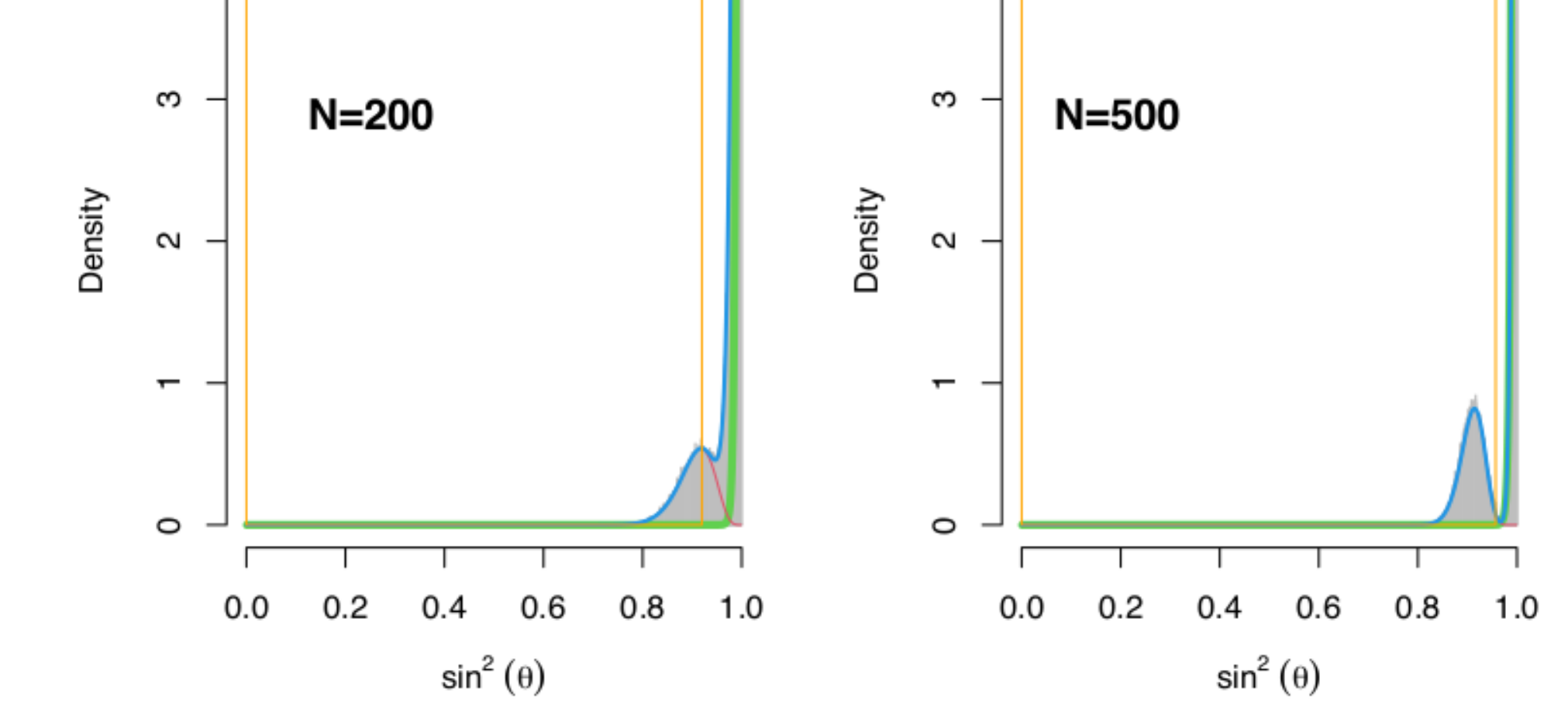}
\caption{Fitted distribution - simulated data: $P=1000$, block-diagonal clustered correlation structure with cluster size 50, $\rho=0.3$.
Left $N=200$, Right $N=500$.}
\label{simfit}
\end{center}
\end{figure}

In Figure \ref{simfit} we show the fitted model for two configurations in the block-diagonal clustered correlation structure with cluster size 50. In both cases, the correlation coefficient was set to $\rho=0.3$. We used  $N=200$ (left) and $N=500$ (right). The histograms in gray show the observed distribution of the $z_j$'s.
The green and red curves represent the fitted null and non-null distributions, respectively. The blue curve depicts the fitted mixture, which fits the data very well. The vertical orange lines show the range in which $z_j$ is deemed small enough so that the corresponding pair is said to be strongly correlated. In this scenario, when $N=200$ the threshold was found to be 0.92, and when $N=500$ any pair with $z_j<0.96$ is declared to be non-null.

\begin{table}[ht]
\centering
\begin{tabular}{rlrrrlrr}
  \hline
 & Corr. Structure & $\rho$ & N & P & Settings & TPR & FDR \\ 
  \hline
  1 & Clusters & 0.3 & 200 & 500 & Cluster size 25 & 0.59 & 1.5e-3 \\ 
  2 & Clusters & 0.9 & 200 & 500 & Cluster size 25 & 1.00 & 4.4e-5 \\ 
  3 & Clusters & 0.3  & 200 & 1000 & Cluster size 100 & 0.66 & 1.2e-3 \\ 
  4 & Clusters & 0.9 & 200 & 1000 & Cluster size 100 & 1.00 & 0.00 \\ 
  5 & Clusters & 0.3 & 200 & 1000 & Cluster size 500 & 0.83 & 5e-4 \\ 
  6 & Clusters & 0.9 & 200 & 1000 & Cluster size 500 & 1.00 & 0.00 \\ 
  7 & Clusters & 0.3 & 500 & 1000 & Cluster size 500 & 0.99 & 1.5e-5 \\ 
  8 & Clusters & 0.9 & 500 & 1000 & Cluster size 500 & 1.00 & 0.00 \\ 
\hline
  9   & Band & 0.3  & 200 & 500 & Width 150 & 0.34 & 5.8e-4 \\ 
  10 & Band & 0.9 & 200 & 500 & Width 150 & 0.92 & 2.7e-3 \\ 
  11 & Band & 0.3  & 200 & 1000 & Width 150 & 0.38 & 1.5e-3 \\ 
  12 & Band & 0.9 & 200 & 1000 & Width 150 & 0.93 & 1.4e-3 \\ 
  13 & Band & 0.3  & 200 & 1000 & Width 30 & 0.38 & 1.7e-3 \\ 
  14 & Band & 0.9 & 200 & 1000 & Width 30 & 0.98 & 5.7e-4 \\ 
  15 & Band & 0.3  & 500 & 1000 & Width 150 & 0.86 & 8.6e-4 \\ 
  16 & Band & 0.9 & 500 & 1000 & Width 150 & 1.00 & 1.6e-3 \\ 
  17 & Band & 0.3  & 500 & 1000 & Width 30 & 0.91 & 3.8e-4 \\ 
  18 & Band & 0.9 & 500 & 1000 & Width 30 & 1.00 & 1.6e-3 \\ 
\hline
  19 & Cycle & 0.3  & 200 & 500 & Length 25 & 0.43 & 4.8e-3 \\ 
  20 & Cycle & 0.9 & 200 & 500 & Length 25 & 1.00 & 2.2e-3 \\ 
  21 & Cycle & 0.3  & 200 & 1000 & Length 50 & 0.32 & 1.9e-3 \\ 
  22 & Cycle & 0.9 & 200 & 1000 & Length 50 & 1.00 & 1.1e-3 \\ 
  23 & Cycle & 0.3  & 500 & 1000 & Length 50 & 0.98 & 1.2e-3 \\ 
  24 & Cycle & 0.9 & 500 & 1000 & Length 50 & 1.00 & 4.7e-3 \\ 
   \hline
\end{tabular}
\caption{Simulation results}\label{simresults}
\end{table}

\section{Data Analysis}
We now demonstrate four applications of the beta-mixture method. 
\subsection{Variable Selection}
\begin{figure}[htbp]
\begin{center}
\includegraphics[scale=0.5]{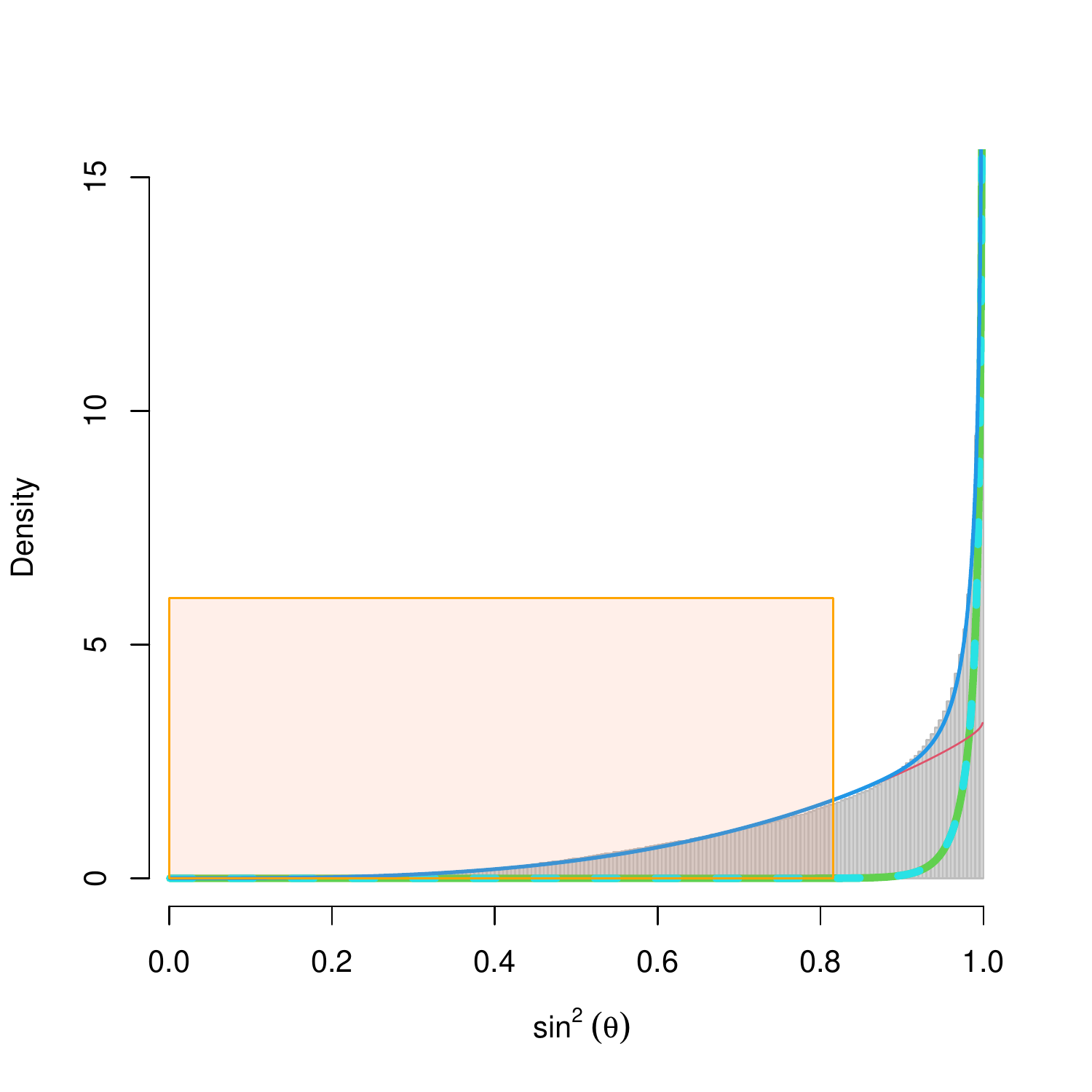}
\caption{The riboflavin data - fitted beta mixture model}
\label{b12fit}
\end{center}
\end{figure}

\begin{figure}[htbp]
\begin{center}
\includegraphics[scale=0.8]{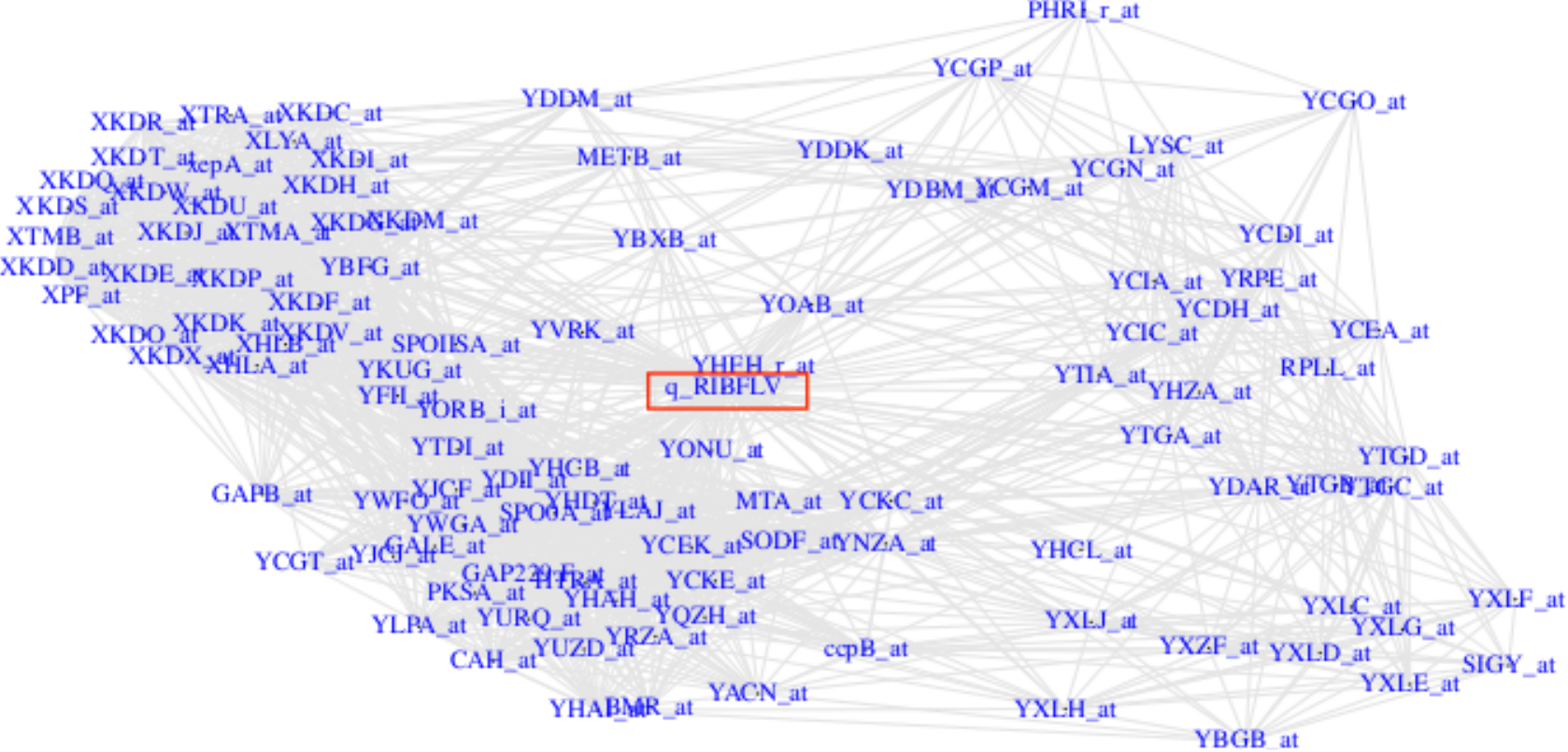}
\caption{106 genes are selected as strong predictors for the production rate of riboflavin data.}
\label{b12links}
\end{center}
\end{figure}

The riboflavin data, which was introduced by \cite{buhlmann2014}, contains normalized expression data of 4,088 genes, and the objective here is to detect which of these genes is a predictor of riboflavin production rate (the response) in \textit{Bacilluss subtilis}. There are $N=71$ samples, which we assume to be independent. Variable selection with the beta-mixture method amounts to detecting the significant correlations or edges between the $P=4088+1$ variables, and ultimately reporting the nodes which are found to be adjacent to the response variable's node.
Figure \ref{b12fit} shows the distribution of the $z_j$'s and the fitted mixture model. The threshold for declaring a pair of variables as significantly correlated is found to be $\sin^2(\theta)>0.815$ ($|r|>0.43$).
For the purpose of variable selection we are only interested in edges which connect to the riboflavin production rate variable (the highlighted node, q\_RIBFLV in Figure \ref{b12links}) and the algorithm selects 106 variables, which form a highly interconnected network with two large clusters of genes.

The large number of selected predictors and the strong dependence among them suggests that riboflavin production is an intricate process which probably cannot be explained satisfactorily by a sparse, linear model. A change in one gene may cause a chain reaction in many other genes, quite possibly involving non-linear effects, thus making it complicated to predict the ultimate effect on the response variable. 

\subsection{Graphical Models}
The tomato seed metabolites profiling data was described and analyzed by \cite{tomatos2017}. One hundred Recombinant Inbred Lines (RILs) were divided into two equal groups -- one was used for dry seeds, and one for six-hour imbibed seeds. The combination of genetic and environmental effects on the metabolic content of tomato seeds is important in order to establish the seed's ability to germinate. In this study, 167 metabolites were detected, of which 68 are known in the literature. Here, we used the beta mixture model to construct and compare the two metabolic networks of the 68 named metabolites.
Among these metabolites, twenty were found to be differentially expressed between the two groups (using t-test, Bonferroni-adjusted p-values less that 0.01, and absolute log fold change greater than 1.) These metabolites are listed in Table \ref{tomatoDEM}.

\begin{figure}[htbp]
\begin{center}
\includegraphics[scale=0.7]{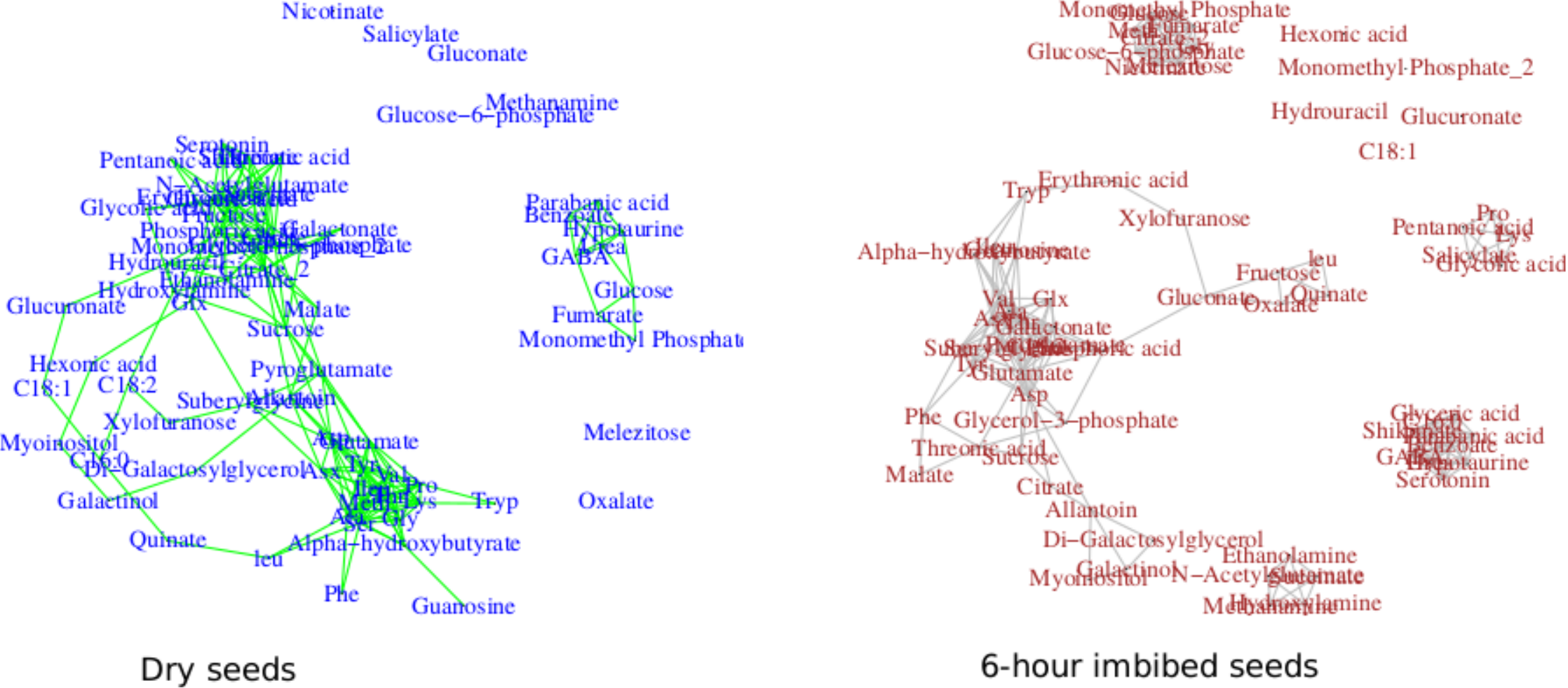}
\caption{Tomato seed metabolic networks -- dry seeds (left) and 6-hour imbibed seeds (right).}
\label{tomatoLinks}
\end{center}
\end{figure}

\begin{figure}[htbp]
\begin{center}
\includegraphics[scale=1]{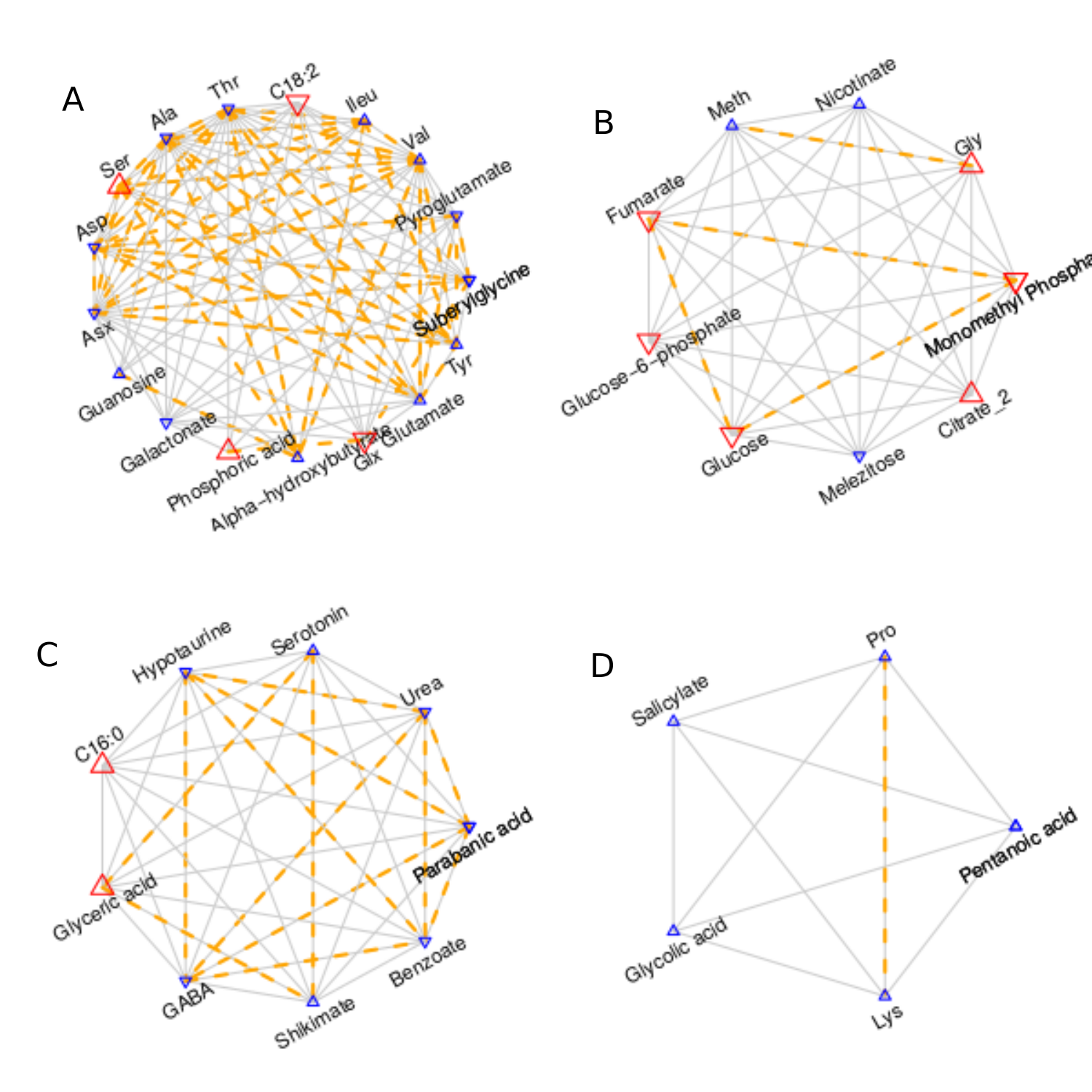}
\caption{Tomato seed metabolic networks -- the four largest clusters in the 6-hour imbibed group. The dashed orange edges represent pairs of metabolites which are also strongly correlated in the dry seed group.}
\label{tomatoClusters}
\end{center}
\end{figure}

The two networks are quite different, as can be seen in Figure \ref{tomatoLinks}. In particular, the six-hour imbibed seeds group has four isolated clusters, all of which form complete graphs. The four largest clusters in the 6-hour imbibed network are shown in Figure \ref{tomatoClusters}. The gray edges represent the strong correlations in the 6-hour imbibed seeds group, and the dashed orange lines represent strong correlations in the dry seeds group. Metabolites are shown as triangles which point up (down) if the expression in the dry seed group is higher (lower) than in the 6-hour imbibed seeds group. A small blue triangle represents a difference which is not significant, and a large red triangle corresponds to one of the differentially expressed metabolites in Table \ref{tomatoDEM}.
The largest cluster (Figure \ref{tomatoClusters} A) contains 17 nodes, forming a near-complete graph. Many of the edges are also significant in the dry seeds group, but there are some potentially meaningful differences. For example, C18:2 (Octadecadienoic acid) which is not part of the dry-seed cluster, and its expression is much higher in the 6-hour imbibed group. Glx (Glutamine) is also more expressed in the 6-hour imbibed group, and it has only two connections in the dry-seed cluster. In this cluster, Phosphoric acid is only connected to Glx in the dry seed network, and its expression in this group is significantly higher. The other three clusters form complete, and disconnected graphs in the 6-hour imbibed group.
Cluster  B in Figure \ref{tomatoClusters} is a complete graph, but only four of the 36 edges also appear in the dry seed graph. Three of the nodes (Glucose, Monomethyl Phosphate, and Fumarate) are significantly more expressed in the 6-hour imbibed group, and they are highly correlated in both groups. Imbibing the seeds not only increases the expression of these metabolites, but also retains their co-expression relationships. Cluster C is also a complete graph, entirely isolated from the rest of the nodes. Its nine nodes consist of the union of two complete and disjoint subgraphs in the dry seed graph (Srotonin + Shikimate + Glyceric acid, and Urea + Parabanic acid + Benzonate + GABA + Hypotaurine), plus C16:0 (Palmitic acid) which is not co-expressed with any of these eight metabolites in the dry seed graph.

We have used the named metabolites in order to provide interpretable results, but using all 167 metabolites may be useful as well, since it may allow us to infer the role of unnamed metabolites based on their association with metabolites with known functionality.

\begin{table}[ht]
\centering
\begin{tabular}{lc}
  \hline
Metabolite & Fold Change \\ 
  \hline
Glucose-6-phosphate & -1.82 \\ 
Glx (Glutamine) & -1.70 \\ 
C18:2 (Octadecadienoic acid) & -1.69 \\ 
Glucose & -1.48 \\ 
Monomethyl Phosphate & -1.35 \\ 
Fumarate & -1.33 \\ 
Gly (Glycine) & 1.18 \\ 
Ser (Serine) & 1.26 \\ 
C18:1 (Oleic acid) & 1.60 \\ 
Fructose & 1.64 \\ 
Threonic acid & 1.96 \\ 
C16:0 (Palmitic acid) & 2.13 \\ 
Phosphoric acid & 3.35 \\ 
Quinate & 3.48 \\ 
leu (Leucine) & 3.65 \\ 
Erythronic acid & 4.70 \\ 
Glyceric acid & 4.75 \\ 
Citrate\_2 & 5.14 \\ 
Gluconate & 5.25 \\ 
Monomethyl Phosphate\_2 & 7.06 \\ 
 \hline
\end{tabular}
\caption{Tomato seed metabolites profiling - differentially expressed metabolites, with $|FC|>1$, $\alpha=0.01$, Bonferroni-adjusted p-values.}\label{tomatoDEM}
\end{table}

\subsection{Spatial Models}\label{spatial}
One of the challenging steps in spatial modeling is to estimate the covariance matrix. Adjacent locations cannot be assumed to be independent, and the spatial correlation must be accounted for. We illustrate how one may use a data-driven approach with the beta mixture model to estimate the spatial covariance matrix (which may depend on covariates, so as to account for differences  between regions.) This can be useful for Kriging of spatial data. We use the  2020 release of the North American Breeding Bird Survey dataset\footnote{https://www.sciencebase.gov/catalog/item/5ea04e9a82cefae35a129d65}, which contains bird species count for more than 700 North American bird taxa. The data is collected each June at thousands of random locations along routes in the United States and Canada. Each route is approximately 40km long, with counting locations placed roughly every 800 meters (50 stops along the route). Counting is performed by a citizen scientist proficient in avian identification.  A longitudinal study could be very interesting in order to detect trends in range, occurrence, and abundance of some birds, and perhaps help to establish ecological health indicators. However, since the number of routes has increased six-fold between 1966 and 2019, and also because conditions may vary significantly between years in some locations and there is only one observation per year, we use just one year (2015) to illustrate our approach.

Our dataset consists of the total number of species count per an entire route. Initially, we have 5,756 locations and 756 unique AOU's (the American Ornithologists' Union identification code for birds). We aggregate counts from routes which are close to each other (within 60km). Birds which have not been observed  in any location, and locations in which no birds have been observed, are eliminated, resulting in $N=608$ birds and $P=601$ locations. The counts are log-transformed in order to normalize the data (adding 1 to all counts, in order to avoid taking the logarithm of zero.)

\begin{figure}[htbp]
\begin{center}
\includegraphics[width=\textwidth]{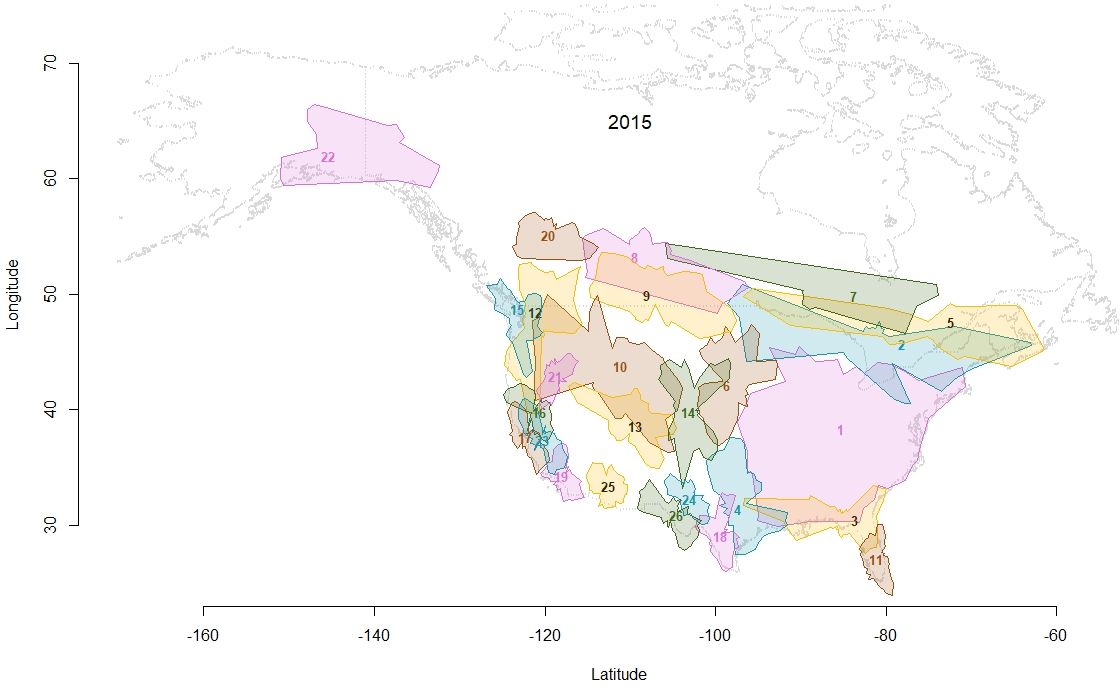}
\caption{The North American Breeding Bird Survey data -- 2015.}
\label{birds2015}
\end{center}
\end{figure}

Since our objective is to obtain a spatial covariance matrix, we treat the locations as our nodes, and use the beta mixture model to find which pairs of locations are strongly correlated. We use the vectors of 608 bird species counts per location to calculate the $z_i$'s which are used when fitting the model. In this type of analysis we must take into account that the observations (bird counts) are not independent, and use the adaptation mentioned in Section \ref{depdata}. This yields an effective sample size $\hat\nu\approx 32$ - much smaller than the actual $N$.
Out of 180,300 possible pairs, 20,349 are detected as highly correlated, so the graph is only relatively sparse (with 11\% of the possible edges).
With these edges, we create 26 clusters (shown in Figure \ref{birds2015}) which consist of locations with similar bird abundance vectors. In the clustering step we first identify the cluster centers based on their degree and clustering coefficient, and then include a location in a cluster if its abundance vector is found to be strongly correlated with the central node by our beta mixture method.
There are many possible clustering methods, and each can be configured with a set of parameters, thus yielding different cluster configurations, but since this is not the focus of this paper we leave the details of our specific choices to be described in the Supplementary Materials.

Some points are worth mentioning about Figure \ref{birds2015}. First, the algorithm uses only bird abundance data, and although the coordinates of points are \textit{not} used in the network construction, the edges found by the beta mixture model yield clusters which correspond very nicely to geographical regions. For example, the Florida panhandle: \#11, the Sonora desert: \#25, along the Missouri River: \#9, \#6, and \#4, and the subarctic region: \#22. The shape and location of the clusters correspond to common habitat conditions, such as climate, vegetation, water resources, and proximity to the shore.
Second, the clusters have a fair amount of overlap (for example, clusters 12, 15, 16, 17, 19, and 23 along the west coast.) The edges we find based on bird abundance correlations allow to capture subtle differences between similar clusters, and although we only have one time point, these overlapping clusters can be used to detect and account for migration paths. Third, notice that some regions are not associated with any cluster. This may very well be due to under-sampling, as is probably the case in the western deserts, and in northern Canada. In spatial data analysis this can be quite important. One may use nearby clusters to impute the covariance structure in such areas, or, possibly infer that the covariance variance is indeed singular in certain regions. For example, a region may be inhabitable, and thus, imposing a non-singular covariance matrix may lead to incorrect predictions.

Obtaining data-driven spatial covariance matrices can be very useful in ecological studies. The approach demonstrated here with one time point from one year may be extended to longitudinal studies, and to consider species-environment interactions and include covariates such as temperature, precipitation, and major events such as hurricanes, wildfires, and volcanic activity.

\subsection{Classification}\label{classification}
To demonstrate how the betaMix approach may be used to perform classification we use the `ionosphere' data \citep{ionosphere1989} from the Machine Learning Repository \citep{Dua2019}. The data was collected from a radar system which aimed radio waves at the ionosphere in order to detect free electrons. If the returned signal does not show evidence of some type of structure, it means that the signal passed through the ionosphere and this radar return should be classified as `bad'. If there is evidence of some type of structure in the radar return then this is classified as good. 
The objective of the original technical report was to show that the process of determining the quality of the return signal can be done automatically, accurately and efficiently by using a multilayer feed forward neural network. At the time of the report the radar in the experiment was producing data every 5 seconds, year round, and since the quality control process was in part manual it was very time-consuming.  

The data has 34 continuous attributes, and a binary outcome -- classification by an expert into `good' or `bad'. Two of the continuous variables have low variability so we exclude them from our analysis (the first variable has 38 zeroes and 313 ones, and the second contains only zeroes). There are 126 bad cases and 225 good ones in the dataset. For the training data we use 60 of each response type, leaving 66 bad and 165 good cases for the test data. As mentioned in Section \ref{sec:method}, we flip the roles of predictors and samples, so instead of considering a matrix with 351 rows and 32 columns, betaMix takes as input a matrix with 32 rows and 351 columns. Our approach is to find the network for the entire data using the betaMix model and the 32 features, and classify each point in the test data based on its similarity (in terms of the adjacency matrix obtained from betaMix) to points in the training data. We use a simple majority rule -- if most of its training set neighbors are good then the point is classified as good. Otherwise it is classified as bad.  To fit the model we set the frequentist error rate parameter to 1e-5 and the Bayesian posterior probability threshold to 0.001. The fitted plot is provided in the Supplementary Material, and the threshold for detecting an edge in the graph is found to be $\sin^2(\theta)<0.56$.
The network for one randomly drawn training data set is shown in Figure \ref{ionosphere}. We use the igraph package \citep{igraph} for this graphical representation.

\begin{figure}[htbp]
\begin{center}
\includegraphics[scale=0.7]{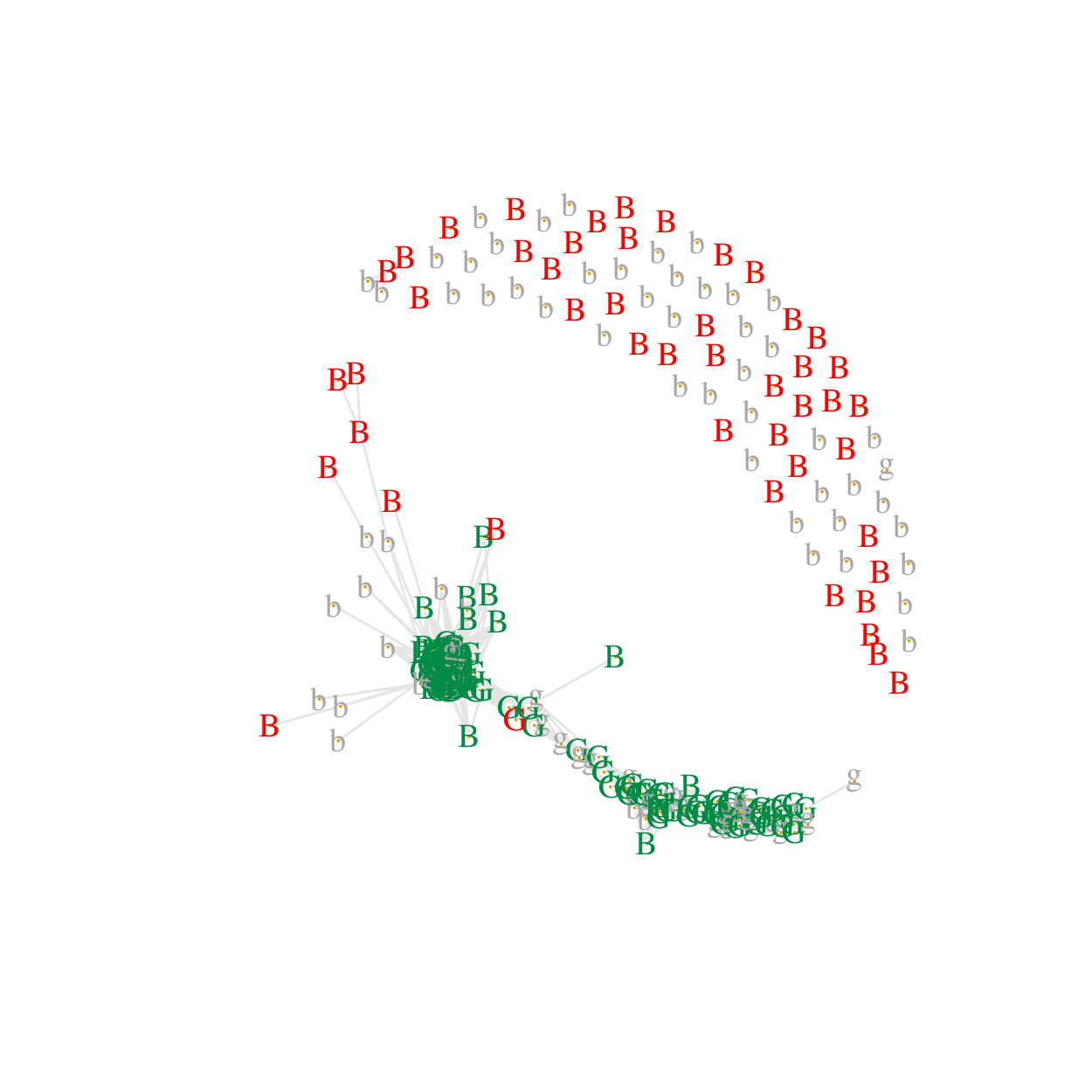}
\caption{The ionosphere network -- lower case letters represent training data, and upper case letters represent test data. The letter specifies the true classification, and the color represents our algorithm's classification (red=bad, green=good). The training dataset has 60 good radar returns and 60 bad ones.}
\label{ionosphere}
\end{center}
\end{figure}

There are a few striking characteristics in the network plot. Most of the good cases are clustered very tightly (in the lower left quadrant). A smaller set of good cases forms a second cluster, with a high degree of connectivity, but less than the main cluster (in the lower right quadrant).  Most of the bad cases are not connected to any other nodes, and among ones which are connected, most appear in the fringes of the clusters because they are similar to a relatively small number of nodes in the cluster. Still, in this particular training data set there are some bad cases which are very similar to the large and tight cluster of good cases. This suggests that correctly classifying all the bad cases is expected to be challenging.

With this typical training data we get an overall accuracy of 91.3\% when we classify the test dataset. The sensitivity is 99.4\% and the specificity is 71.2\% (correctly classifying 47 of the 66 bad cases). From the plot we see that we can improve the results by using a stricter condition for the classification rule. For example, if we only classify a point as good if it has at least four training set neighbors, of which the majority are good, then the accuracy increases to 94.8\%, the sensitivity is 98.7\% and the specificity is 84.8\%.
To improve the specificity further we may consider using higher orders of the 32 variables in the input to betaMix. However, this is outside the scope of this article. Our goal here is to demonstrate that networks obtained from the betaMix model can be used to perform accurate classification, while also providing insights about the relationships among samples and classes.
An additional example involving Congress voting data is provided in the Supplementary Material.


\section{Discussion}\label{sec:disc}
We have introduced a mixture-model of beta distributions to identify significant correlations among $P$ predictors when $P$ is large. The method relies on theorems in convex geometry, which were used here to show how to control the error rate of edge detection in graphical models.
The betaMix method does not require any assumptions about the network structure, nor does it assume that the network is sparse. When the network is dense the null probability parameter may be estimated as $\hat{p}_0=0$ and a single beta component will be used to fit the data. However, in the applications discussed here, such as co-expression of genes or metabolites or co-abundance of birds in spatial modeling, it is expected that many pairs (genes, metabolites, locations) will be uncorrelated, and betaMix has been motivated by such applications. Applying betaMix to datasets in which there are no uncorrelated pairs of nodes in the graph is possible, but may require some modifications to the model. For example, it may be the case that while there are no uncorrelated pairs of nodes, there are still many `mostly uncorrelated' ones. So, from the modeling perspective the null component may be allowed to be concentrated near 1, but not with $(N-1)/2$ and $1/2$ parameters as implied from the convex geometry theory. This will require a different geometric/probabilistic justification. Similarly, the correlated pairs may actually arise from two or more regimes, in which case it may be better for the non-null component to be a mixture of multiple beta distributions. These extensions are left for future research.

Another future avenue of research is to explore the possibility to extend betaMix for the analysis of causal models. In such models the edges need to be directional, while the method presented here only accommodates bi-directional ones, since it relies on correlations which are symmetric. 

As mentioned in section \ref{spatial}, it will be useful to incorporate covariates in the estimation of edges, since the existence of an edge in the graph may depend on time, location, temperature, and so on. It will also be interesting to explore the possibility to extend the approach beyond linear correlations. For example, the association between variables may be strong only when considering their upper (or lower) quantiles, but they may not be correlated in the usual sense (Pearson or Spearman).

The R package which implements the beta-mixture model is available from github, at \url{https://github.com/haimbar/betaMix} and data and code files used in this paper can be found at \url{https://github.com/haimbar/betaMixFiles}.


\bigskip


\appendix
\section{Appendix}\label{sec:app}
\subsection{Likelihood Expressions}
The log-likelihood function derived from the mixture model is  
\begin{eqnarray*}
\ell(\mathbf{z})&=& \sum_{j=1}^M\left[ m_{0j}\frac{1}{B(\frac{\nu-1}{2},\frac{1}{2})}z_j^{\frac{\nu-1}{2}-1}(1-z_j)^{-\frac{1}{2}}+(1-m_{0j})\frac{1}{B(a,b)}z_j^{a-1}(1-z_j)^{b-1}\right]
\end{eqnarray*}
where $m_{0j}\sim Ber(p_0)$ and $M$ is the total number of possible edges. The maximum likelihood estimates of $a$, $b$, and $\nu$ are obtained by solving the following equations, with $\psi()$ denoting the digamma function:
\begin{eqnarray}
\psi(a)-\psi(a+b) &=& \frac{\sum_{j=1}^M(1-m_{0j})\log(z_j)}{\sum_{j=1}^M(1-m_{0j})}\\
\psi(b)-\psi(a+b) &=& \frac{\sum_{j=1}^M(1-m_{0j})\log(1-z_j)}{\sum_{j=1}^M(1-m_{0j})}\\
\psi((\nu-1)/2)-\psi(\nu/2) &=& \frac{\sum_{j=1}^Mm_{0j}\log(z_j)}{\sum_{j=1}^M m_{0j}}\,.
\end{eqnarray}
The estimates for the Bernoulli variables are obtained by plugging in the estimated expected values under the mixture model:
\begin{eqnarray}
\hat{m}_{0j}&=&\frac{\hat{p}_0\hat{f}_0(z_j)}{\hat{p}_0\hat{f}_0(z_j)+(1-\hat{p}_0)\hat{f}(z_j)}\,,
\end{eqnarray}
and $\hat{p}_0=\frac{1}{M}\sum_{j=1}^Mm_{0j}$.

\subsection{Gauss Hypergeometric Functions}
Many common special functions in that arise in statistics are particular cases of the \textit{Gauss
hypergeometric series} defined by

\begin{equation}\label{hyp1}
_2 F _1 (a,b; c; z) = \sum_{k=0}^\infty \frac{(a)_k (b)_k}{(c)_k}
\frac{z^k}{k!} ,
\end{equation}
where the rising factorial $(a)_k$ is defined by $(a)_0 = 1$ and

\begin{equation*}
(a)_k = a(a+1) \cdots (a+k-1), \quad (k \ge 1) ,
\end{equation*}

\noindent for arbitrary $a \in \mathbb{R}$.  The series (\ref{hyp1}) is not defined when
$c = - m$, with $m=0, 1, 2, \ldots$, unless $a$ or $b$ are equal to
$- n$, $n = 0, 1, 2, \ldots$, and $n < m$.  It is also easy to see
that the series (\ref{hyp1}) reduces to a polynomial of degree $n$ in
$z$ when $a$ or $b$ is equal to $- n$, $n=0, 1, 2, \ldots$.  In all
other cases the series has radius of convergence 1. The hypergeometric function also has the integral representation 
\begin{equation}\label{beta3}
{_2F_1}{\left(a,b;c;z\right)}=\frac{\Gamma(c)}{\Gamma(a)\Gamma(c-b)}\int_{0}^{1}\,\frac{t^{a-1}\left(1-t\right)^{c-a-1}}{\left(1-zt\right)^{b}} dt
\end {equation}
where $0<t<1$, $b>0$, and $c>a$.

The function defined by the series (\ref{hyp1}) and integral (\ref{beta3}) is called the {\it Gauss hypergeometric function}.  Many properties of the Gauss hypergeometric function can be found in the classic reference works by  \cite{erdelyi} and \cite{magnus}.

Many distribution of random matrices can be expressed in terms of a Gauss hypergeometric functions of a matrix argument.  A detailed introduction to hypergeometric functions of a matrix argument can be found in \cite{muirhead1982}.  Their basic definition depends on zonal polynomials which are defined in terms of partitions of positive integers and a matrix valued extension of the series in (\ref{hyp1}).  Another representation of the Gauss hypergeometric functions of a matrix argument is in terms of the integral

\begin{eqnarray}\label{hyper3}
{_2F_1}(a,b;c;Z)&=&\frac{\Gamma_m(c)}{\Gamma_m(a)\Gamma_m(c-a)}\int_{0<Y<I_m}\,\det(I-ZY)^{-b} (\det Y)^ {a-(m+1)/2} \\ \nonumber
& & \quad \cdot \det(I-Y)^{c-a-(m+1)/2} \; (dY).
\end{eqnarray}
for $Re(Z)<I$, $b>(m-1)/2, c-a>(m-1)/2$ and where 
$\Gamma_m(a) =\pi^{m(m-1)/4} \prod_{i=1}^{m} \Gamma[a-\frac{1}{2}(i-1)]$
is the multivariate gamma function.  \citet{koev2006efficient} have developed efficient algorithms to evaluate ${_2F_1}$ with matrix argument, the software implementation can be found in \citet{koev2008hypergeometric} in Matlab and {\tt Hypergeom2F1Mat.R} in the R library CharFunToolR \citep{CharFunToolR}.

\bibliographystyle{chicago}
\bibliography{BetaSelectionArxiv}

\clearpage

\begin{center}
\section*{Supplementary Material}
\end{center}
\beginsupplement

\section*{Complete Simulation Results}
\begin{itemize}
 \item Scenario \#1 (Table \ref{sim01}) -- a simple linear model. $X$ is an $N\times P$ matrix, drawn from a uniform distribution, and $Y=1.6+6X_1+4X_{30}+3X_{100}+\epsilon$ where the error term is $N(0,0.1^2)$. This is a simulation for testing the ability of betaMix to recover the true model in high-dimensional $\beta$-sparse setting.
 \item Scenario \#2 (Table \ref{sim02}) -- Similar to the previous scenario, except that the predictors are not independent. The first 15 predictors are correlated with either an AR(1) structure, or a hub structure where $X_1$ being the center of the hub. This is a simulation for testing the ability of betaMix to recover the true model in high-dimensional $\beta$-sparse setting, in the presence of varying degrees of correlations among the predictors.
 \item Scenario \#3 (Table \ref{sim03}) -- the $P\times P$ covariance matrix has an AR(1) structure with correlation coffined $\rho$.
 \item Scenario \#4 (Table \ref{sim04}) -- similar to the previous scenario, but only the first block of $K=20$ variables has an AR(1) correlation structure, and the rest are i.i.d.
 \item Scenario \#5 (Table \ref{sim05}) -- the $P\times P$ covariance matrix has a band structure with correlation coffined $\rho$ and band-width $BW$.
 \item Scenario \#6 (Table \ref{sim06}) -- the $P\times P$ covariance matrix has a block-diagonal cluster structure with correlation coffined $\rho$ and varying cluster sizes $C$.
 \item Scenario \#7 (Table \ref{sim07}) -- similar to \#6, with 40 clusters, but each with a different size.
 \item Scenario \#8 (Table \ref{sim08}) -- similar to \#6, except that the blocks are hubs of size $H$, not complete clusters.
 \item Scenario \#9 (Table \ref{sim09}) -- Cycle configurations - each cluster of size $S$ forms a cycle, e.g. 1--2--3--...--S--1
\end{itemize}

\begin{table}[h!]
\centering
\begin{tabular}{ccccc}
  \hline
 N & P &  TPR & FDR \\ 
  \hline
 200 &  500 & 0.956 & 0.000 \\ 
 200 & 1000 & 0.944 & 0.000 \\ 
 500 &  500 & 1.000 & 0.000 \\ 
 500 & 1000 & 1.000 & 0.000 \\ 
   \hline
\end{tabular}
\caption{Simulation results - simple linear model}\label{sim01}
\end{table}

\begin{table}[h!]
\centering
\begin{tabular}{ccccc}
  \hline
    N & P & Structure, $\rho$ & TPR & FDR \\ 
  \hline
  200 &  500 & AR(1) 0.3 & 0.171 & 0.000 \\ 
  200 &  500 & AR(1) 0.9 & 0.953 & 0.000 \\ 
  200 & 1000 & AR(1) 0.3 & 0.157 & 0.000 \\ 
  200 & 1000 & AR(1) 0.9 & 0.951 & 0.000 \\ 
  500 &  500 & AR(1) 0.3 & 0.220 & 0.000 \\ 
  500 &  500 & AR(1) 0.9 & 1.000 & 0.000 \\ 
  500 & 1000 & AR(1) 0.3 & 0.227 & 0.000 \\ 
  500 & 1000 & AR(1) 0.9 & 1.000 & 0.000 \\ 
  \hline
  200 &  500 & hub 0.3 & 0.261 & 0.006 \\ 
  200 &  500 & hub 0.9 & 0.986 & 0.000 \\ 
  200 & 1000 & hub 0.3 & 0.231 & 0.000 \\ 
  200 & 1000 & hub 0.9 & 0.976 & 0.000 \\ 
  500 &  500 & hub 0.3 & 0.810 & 0.000 \\ 
  500 &  500 & hub 0.9 & 1.000 & 0.002 \\ 
  500 & 1000 & hub 0.3 & 0.708 & 0.000 \\ 
  500 & 1000 & hub 0.9 & 1.000 & 0.000 \\ 
   \hline
\end{tabular}
\caption{Simulation results - sparse linear model, with correlated predictors.}\label{sim02}
\end{table}

\begin{table}[h!]
\centering
\begin{tabular}{ccccc}
  \hline
 N & P & $\rho$ & TPR & FDR \\ 
  \hline
  200 &  500 & 0.3 & 0.425 & 0.000 \\ 
  200 &  500 & 0.9 & 1.000 & 0.000 \\ 
  200 & 1000 & 0.3 & 0.309 & 0.000 \\ 
  200 & 1000 & 0.9 & 1.000 & 0.000 \\ 
  500 &  500 & 0.3 & 0.991 & 0.000 \\ 
  500 &  500 & 0.9 & 1.000 & 0.000 \\ 
  500 & 1000 & 0.3 & 0.983 & 0.000 \\ 
  500 & 1000 & 0.9 & 1.000 & 0.000 \\ 
   \hline
\end{tabular}
\caption{Simulation results - AR(1)}\label{sim03}
\end{table}

\begin{table}[h!]
\centering
\begin{tabular}{ccccc}
  \hline
 N & P & $\rho$ & TPR & FDR \\ 
  \hline
  200 &  500 & 0.3 & 0.446 & 0.000 \\ 
  200 &  500 & 0.9 & 1.000 & 0.000 \\ 
  200 & 1000 & 0.3 & 0.314 & 0.000 \\ 
  200 & 1000 & 0.9 & 1.000 & 0.000 \\ 
  500 &  500 & 0.3 & 0.993 & 0.000 \\ 
  500 &  500 & 0.9 & 1.000 & 0.000 \\ 
  500 & 1000 & 0.3 & 0.989 & 0.000 \\ 
  500 & 1000 & 0.9 & 1.000 & 0.000 \\ 
   \hline
\end{tabular}
\caption{Simulation results - block AR(1)} \label{sim04}
\end{table}

\begin{table}[h!]
\centering
\begin{tabular}{ccccc}
  \hline
 N & P & $BW, \rho$ & TPR & FDR \\ 
  \hline
  200 &  500 & 30, 0.3 & 0.453 & 0.001 \\ 
  200 &  500 & 30, 0.9 & 0.981 & 0.001 \\ 
  200 & 1000 & 30, 0.3 & 0.384 & 0.002 \\ 
  200 & 1000 & 30, 0.9 & 0.976 & 0.001 \\ 
  500 &  500 & 30, 0.3 & 0.930 & 0.000 \\ 
  500 &  500 & 30, 0.9 & 1.000 & 0.002 \\ 
  500 & 1000 & 30, 0.3 & 0.915 & 0.000 \\ 
  500 & 1000 & 30, 0.9 & 1.000 & 0.002 \\ 
  \hline
  200 &  500 & 150, 0.3 & 0.345 & 0.001 \\ 
  200 &  500 & 150, 0.9 & 0.924 & 0.003 \\ 
  200 & 1000 & 150, 0.3 & 0.381 & 0.001 \\ 
  200 & 1000 & 150, 0.9 & 0.933 & 0.001 \\ 
  500 &  500 & 150, 0.3 & 0.870 & 0.000 \\ 
  500 &  500 & 150, 0.9 & 0.993 & 0.002 \\ 
  500 & 1000 & 150, 0.3 & 0.865 & 0.001 \\ 
  500 & 1000 & 150, 0.9 & 0.999 & 0.002 \\ 
   \hline
\end{tabular}
\caption{Simulation results - Band}\label{sim05}
\end{table}

\begin{table}[h!]
\centering
\begin{tabular}{ccccc}
  \hline
 N & P & $C, \rho$ & TPR & FDR \\ 
  \hline
  200 &  500 & 25, 0.3 & 0.588 & 0.001 \\ 
  200 &  500 & 25, 0.9 & 1.000 & 0.000 \\ 
  500 &  500 & 25, 0.3 & 0.991 & 0.000 \\ 
  500 &  500 & 25, 0.9 & 1.000 & 0.000 \\ 
  \hline
  200 &  500 & 50, 0.3 & 0.658 & 0.001 \\ 
  200 &  500 & 50, 0.9 & 1.000 & 0.000 \\ 
  200 & 1000 & 50, 0.3 & 0.594 & 0.001 \\ 
  200 & 1000 & 50, 0.9 & 1.000 & 0.000 \\ 
  500 &  500 & 50, 0.3 & 0.992 & 0.000 \\ 
  500 &  500 & 50, 0.9 & 1.000 & 0.000 \\ 
  500 & 1000 & 50, 0.3 & 0.987 & 0.000 \\ 
  500 & 1000 & 50, 0.9 & 1.000 & 0.000 \\ 
  \hline
  200 & 1000 & 100, 0.3 & 0.655 & 0.001 \\ 
  200 & 1000 & 100, 0.9 & 1.000 & 0.000 \\ 
  500 & 1000 & 100, 0.3 & 0.990 & 0.000 \\ 
  500 & 1000 & 100, 0.9 & 1.000 & 0.000 \\ 
  \hline
  200 &  500 & 250, 0.3 & 0.793 & 0.000 \\ 
  200 &  500 & 250, 0.9 & 1.000 & 0.000 \\ 
  500 &  500 & 250, 0.3 & 0.995 & 0.000 \\ 
  500 &  500 & 250, 0.9 & 1.000 & 0.000 \\ 
  \hline
  200 & 1000 & 500, 0.3 & 0.828 & 0.000 \\ 
  200 & 1000 & 500, 0.9 & 1.000 & 0.000 \\ 
  500 & 1000 & 500, 0.3 & 0.995 & 0.000 \\ 
  500 & 1000 & 500, 0.9 & 1.000 & 0.000 \\ 
   \hline
\end{tabular}
\caption{Simulation results - clusters}\label{sim06}
\end{table}

\begin{table}[h!]
\centering
\begin{tabular}{ccccc}
  \hline
 N & P & $\rho$ & TPR & FDR \\ 
  \hline
  200 &  500 & 0.3 & 0.588 & 0.001 \\ 
  200 &  500 & 0.9 & 1.000 & 0.000 \\ 
  200 & 1000 & 0.3 & 0.590 & 0.001 \\ 
  200 & 1000 & 0.9 & 1.000 & 0.000 \\ 
  500 &  500 & 0.3 & 0.991 & 0.000 \\ 
  500 &  500 & 0.9 & 1.000 & 0.000 \\ 
  500 & 1000 & 0.3 & 0.983 & 0.000 \\ 
  500 & 1000 & 0.9 & 1.000 & 0.000 \\ 
   \hline
\end{tabular}
\caption{Simulation results - random size clusters}\label{sim07}
\end{table}

\begin{table}[h!]
\centering
\begin{tabular}{ccccc}
  \hline
 N & P & $H, \rho$ & TPR & FDR \\ 
  \hline
  200 &  500 & 10, 0.3 & 0.710 & 0.003 \\ 
  200 &  500 & 10, 0.9 & 0.975 & 0.004 \\ 
  500 &  500 & 10, 0.3 & 1.000 & 0.002 \\ 
  500 &  500 & 10, 0.9 & 1.000 & 0.014 \\ 
  \hline
  200 & 1000 & 20, 0.3 & 0.129 & 0.008 \\ 
  200 & 1000 & 20, 0.9 & 0.596 & 0.002 \\ 
  500 & 1000 & 20, 0.3 & 0.851 & 0.001 \\ 
  500 & 1000 & 20, 0.9 & 0.999 & 0.005 \\ 
  \hline
  200 &  500 & 25, 0.3 & 0.118 & 0.017 \\ 
  200 &  500 & 25, 0.9 & 0.556 & 0.004 \\ 
  500 &  500 & 25, 0.3 & 0.797 & 0.002 \\ 
  500 &  500 & 25, 0.9 & 0.998 & 0.007 \\ 
  \hline
  200 & 1000 & 50, 0.3 & 0.011 & 0.078 \\ 
  200 & 1000 & 50, 0.9 & 0.148 & 0.007 \\ 
  500 & 1000 & 50, 0.3 & 0.225 & 0.003 \\ 
  500 & 1000 & 50, 0.9 & 0.868 & 0.003 \\ 
   \hline
\end{tabular}
\caption{Simulation results - hub}\label{sim08} 
\end{table}

\begin{table}[h!]
\centering
\begin{tabular}{cccccc}
  \hline
  N & P & $S, \rho$ & TPR & FDR \\ 
  \hline
  200 &  500 & 25, 0.3 & 0.435 & 0.005 \\ 
  200 &  500 & 25, 0.9 & 1.000 & 0.002 \\ 
  500 &  500 & 25, 0.3 & 0.990 & 0.002 \\ 
  500 &  500 & 25, 0.9 & 1.000 & 0.009 \\ 
  \hline
  200 &  500 & 50, 0.3 & 0.429 & 0.003 \\ 
  200 &  500 & 50, 0.9 & 1.000 & 0.003 \\ 
  200 & 1000 & 50, 0.3 & 0.320 & 0.002 \\ 
  200 & 1000 & 50, 0.9 & 1.000 & 0.001 \\ 
  500 &  500 & 50, 0.3 & 0.993 & 0.002 \\ 
  500 &  500 & 50, 0.9 & 1.000 & 0.010 \\ 
  500 & 1000 & 50, 0.3 & 0.983 & 0.001 \\ 
  500 & 1000 & 50, 0.9 & 1.000 & 0.005 \\ 
  \hline
  200 & 1000 & 100, 0.3 & 0.325 & 0.002 \\ 
  200 & 1000 & 100, 0.9 & 1.000 & 0.001 \\ 
  500 & 1000 & 100, 0.3 & 0.983 & 0.001 \\ 
  500 & 1000 & 100, 0.9 & 1.000 & 0.004 \\ 
  \hline
  200 &  500 & 250, 0.3 & 0.434 & 0.004 \\ 
  200 &  500 & 250, 0.9 & 1.000 & 0.002 \\ 
  500 &  500 & 250, 0.3 & 0.993 & 0.002 \\ 
  500 &  500 & 250, 0.9 & 1.000 & 0.009 \\ 
  \hline
  200 & 1000 & 500, 0.3 & 0.314 & 0.003 \\ 
  200 & 1000 & 500, 0.9 & 1.000 & 0.001 \\ 
  500 & 1000 & 500, 0.3 & 0.983 & 0.001 \\ 
  500 & 1000 & 500, 0.9 & 1.000 & 0.004 \\ 
   \hline
\end{tabular}
\caption{Simulation results - cycle configuration}\label{sim09} 
\end{table}

\pagebreak
\clearpage

\section*{Data Analysis}
\subsection*{Bird abundance data}
\begin{figure}[h!]
\begin{center}
\includegraphics[width=0.6\textwidth]{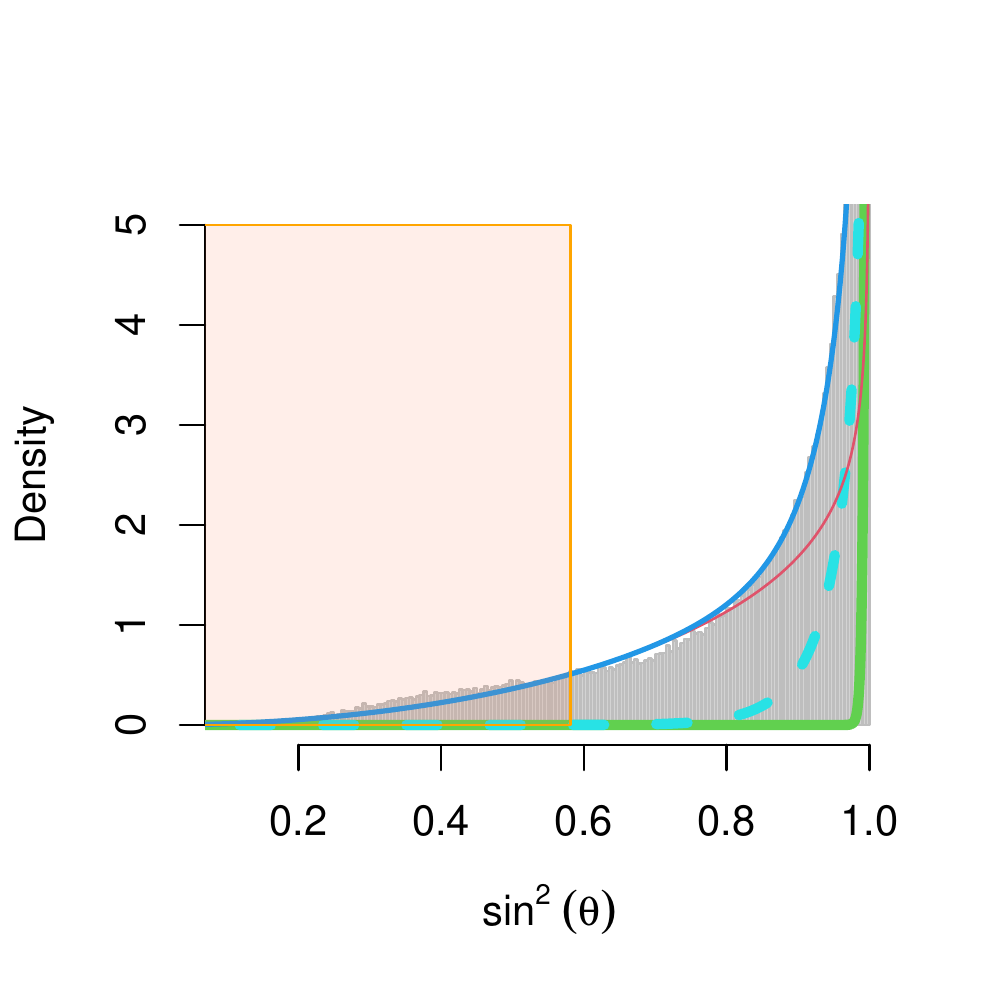}
\caption{The fitted beta mixture model for the North American Breeding Bird Survey data, 2015.}
\label{birds2015fitted}
\end{center}
\end{figure}

Figure \ref{birds2015fitted} shows the histogram of $\sin^2(\theta)$ for the bird abundance data from 2015 and the fitted distribution. The solid green line is the distribution of the null component under the assumption of independent samples. The dashed line is the distribution of the null component if we do not assume independence, and instead estimate the effective sample size. In this case, there is no reason to assume that the counts in different locations are independent. Neighboring locations are expected to have similar species abundance vectors. The red curve is the fitted distribution of the non-null component. The orange region shows The range of $\sin^2(\theta)$ for which pairs are considered highly correlated.

\begin{figure}[h!]
\begin{center}
\includegraphics[width=\textwidth]{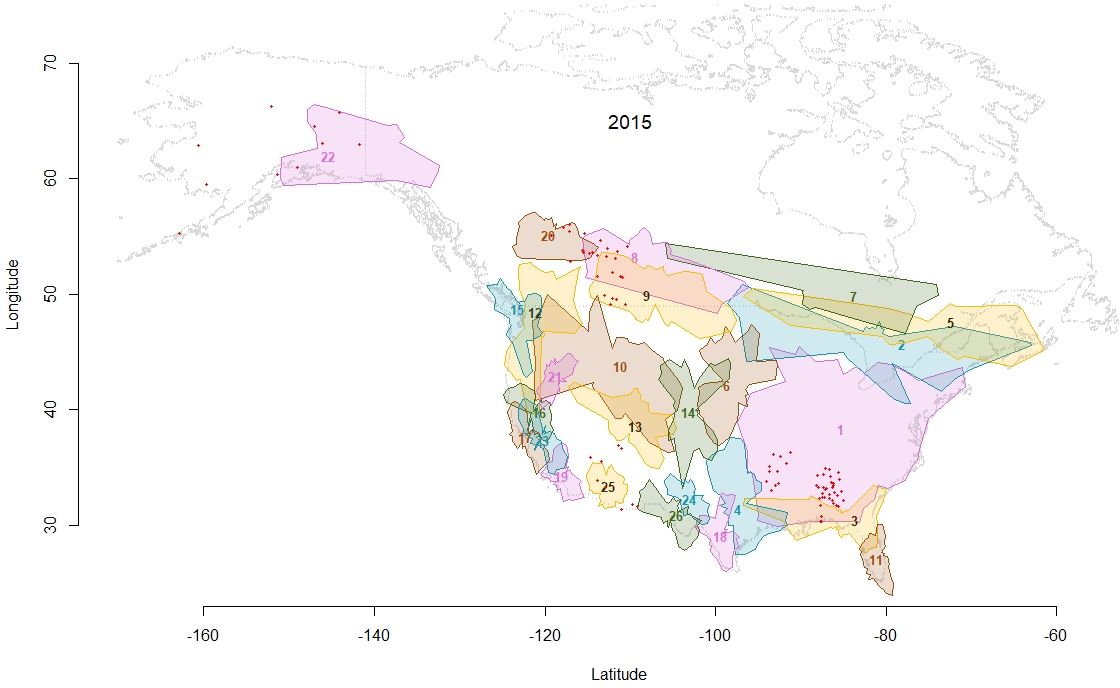}
\caption{The North American Breeding Bird Survey data -- 2015.}
\label{birds2015b}
\end{center}
\end{figure}

With the edges detected by the betaMix algorithm we find 26 clusters, shown in Figure \ref{birds2015b}. Note that there are 100 locations which were not assigned to any cluster (the red dots). As mentioned in the paper, for the clustering step we calculate  each node's degree, $d_j$, and clustering coefficient, $c_j$. In this case we define the centrality by $k_j=d_jc_j$. We then find the node with the largest $k_{j_1}$ and define it as the central node in the first cluster, which contains all the nodes which are connected to it via an edge in the graph. Then, among the unclustered nodes we find the one with the largest $k_{j_2}$ and make it the center of cluster \#2, which contains all the nodes which are connected to node $j_2$. We continue this way, but since we require that $k_j$ must be at least 3, the algorithm may end before all the nodes are assigned to a cluster. The unassigned nodes may be very similar to nodes in an existing cluster, but if there is no edge between a node $x$ to the central node, then $x$ will not be assigned to that cluster.
There are many other clustering methods, and each can yield different configuration. For example, we could have defined $k_j=d_j$, or we could have used an agglomerative clustering method to make sure that all the nodes are assigned to some cluster. However, having unclustered nodes may be interesting by itself. For example, some nodes in the more remote corners of Alaska, may indeed be different in their species abundance than the closest cluster (\#22). Or, the points which lie in the boundary between clusters 1 and 3 may be similar to both, but sufficiently different to be left out of either.

\subsection*{The ionosphere radar data}
Figure \ref{ionosphere} shows the histogram of $\sin^2(\theta)$ for the radar data from the Machine Learning Repository\footnote{https://archive.ics.uci.edu/ml/datasets/ionosphere} and the fitted beta mixture distribution.

\begin{figure}[h!]
\begin{center}
\includegraphics[scale=0.6]{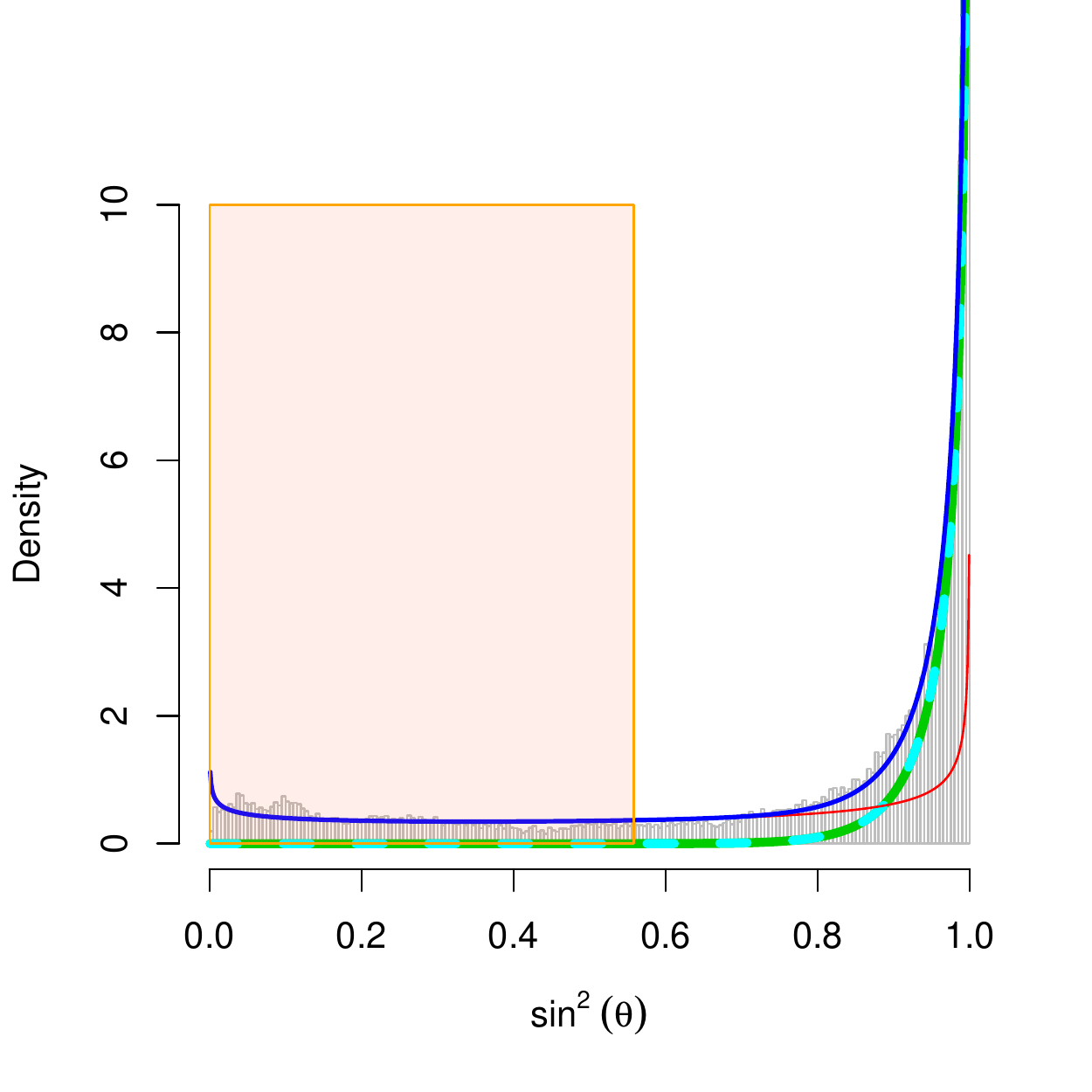}
\caption{The histogram of $\sin^2(\theta)$ from the ionosphere radar data and the fitted betaMix distribution.}
\label{ionosphere}
\end{center}
\end{figure}

\subsection*{Congress voting data}
The American political system is considered highly polarized, with the two major parties disagreeing on most issues. We obtained voting records from the Senate and the House of Representatives from 1993 to 2019, which includes the tenure of four presidents -- William J. Clinton (1993-2000), George W. Bush (2001-2008), Barack H. Obama (2009-2016), and Donald J. Trump (2016-2020). The data consists of 14 Congress election cycles (from the 103rd to the 116th Congress).
In the following, we use the voting data from 2008.

Our general approach is to perform unsupervised clustering using the voting history as characteristics of representatives to find similarities between members of Congress. To do that, we consider each representative as a node in graph, and our first objective is to find which pairs have similar voting patterns in a given year and connect such pairs of nodes with an edge. Then, using the obtained graph we compute useful node characteristics, such as the degree and clustering coefficient, and find clusters in the graph. The algorithm yields two large clusters separating the two major parties (Democratic and Republican).

Congress voting data is publicly available from\\ \url{https://www.govtrack.us/congress/votes}, which contains the following description of the data: 
\begin{quote}
Each year the U.S. Senate and House of Representatives take thousands of votes, some to pass bills, resolutions, nominations, and treaties, and others on procedural matters such as on cloture and other motions. Not all votes are recorded, such as when there is no one opposed. This page shows the outcome of all recorded votes on the Senate floor and House floor. It does not include votes in committee.
\end{quote}
A proposed resolution is given an issue number, and it may correspond to several votes (`roll calls'). Since votes on the same issue are highly correlated we use one roll call for an issue. We exclude any issue for which the vote was unanimous, and exclude members who have participated in less than 30\% of the roll calls in that year. Voting records are stored in a matrix (denoted by $M$) in which rows correspond to members of Congress, and columns to issues. Cell $i,j$ in the matrix contains 0.5, -0.5, or 0 if the $i$-th member voted in favor of the $j$-th bill, against it, or was marked as `Not voting' or `Present', respectively.
To infer the Congress network structure based on voting patterns we look for strong correlations between pairs of rows in $M$. We denote the total number of members (nodes in the graph) by $P$, and the total number of issues by $N$, so $M$ is an $P\times N$ matrix. We use the betaMix model to find the edges between congress members. In this example, $N=332$ and $P=433$.  The maximum number of edges in the graph is 93,528 and betaMix finds 40,320 edges in the graph (using the criterion $\sin^2(\theta)< 0.72$). With 43\% of the possible edges present, the graph is not sparse at all.

\begin{figure}[t!]
\begin{center}
\includegraphics[scale=0.6]{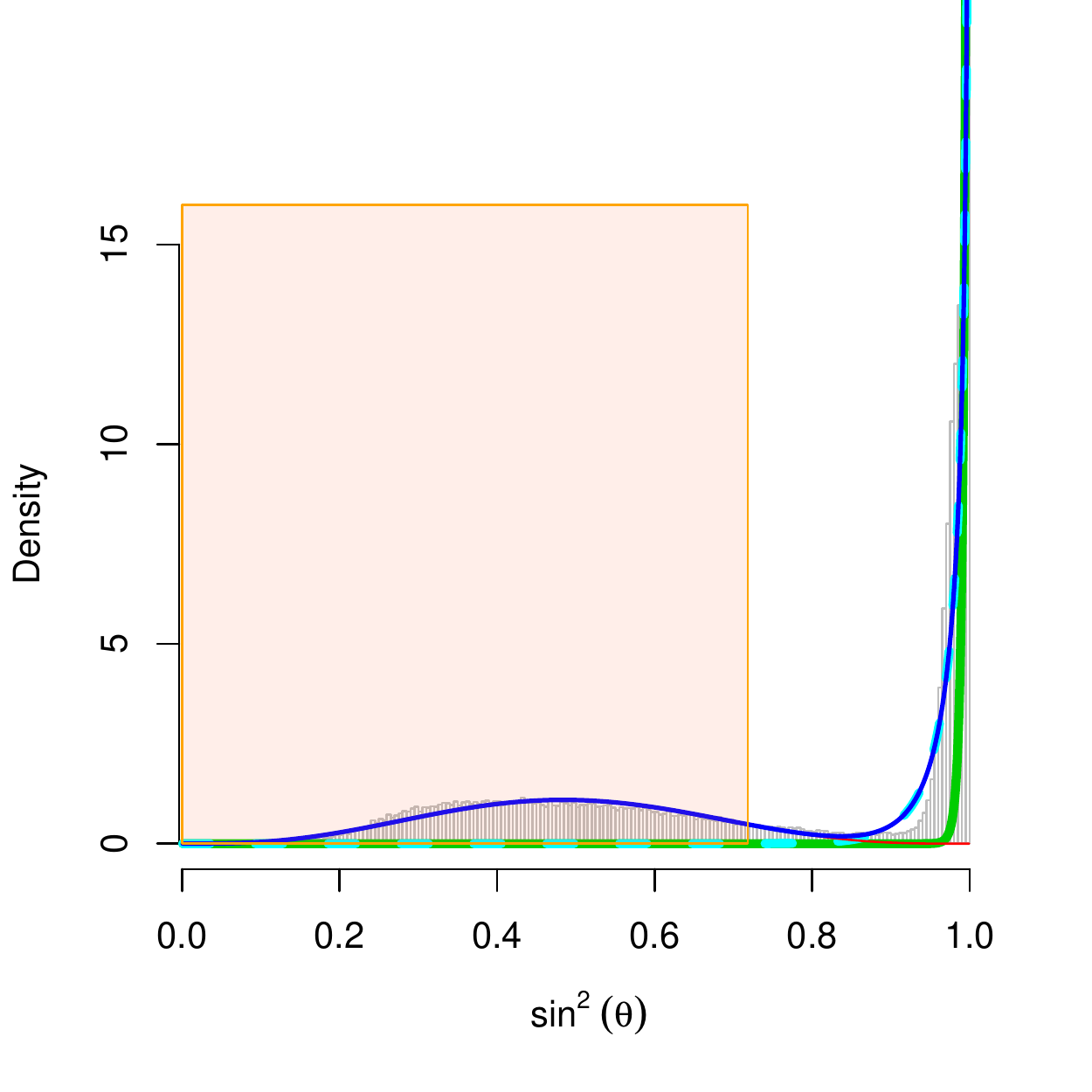}
\caption{The histogram of $\sin^2(\theta)$ from the Congress voting data for 2008 and the fitted betaMix distribution.}
\label{congress2008}
\end{center}
\end{figure}

Figure \ref{congress2008} shows the histogram of $\sin^2(\theta)$ from the Congress voting data for 2008 and the fitted betaMix distribution. Note that there is no reason to assume independence in this case, and the estimated effective sample size is approximately 47 - much smaller than $N$.

\begin{figure}[b!]
\begin{center}
\includegraphics[scale=0.8]{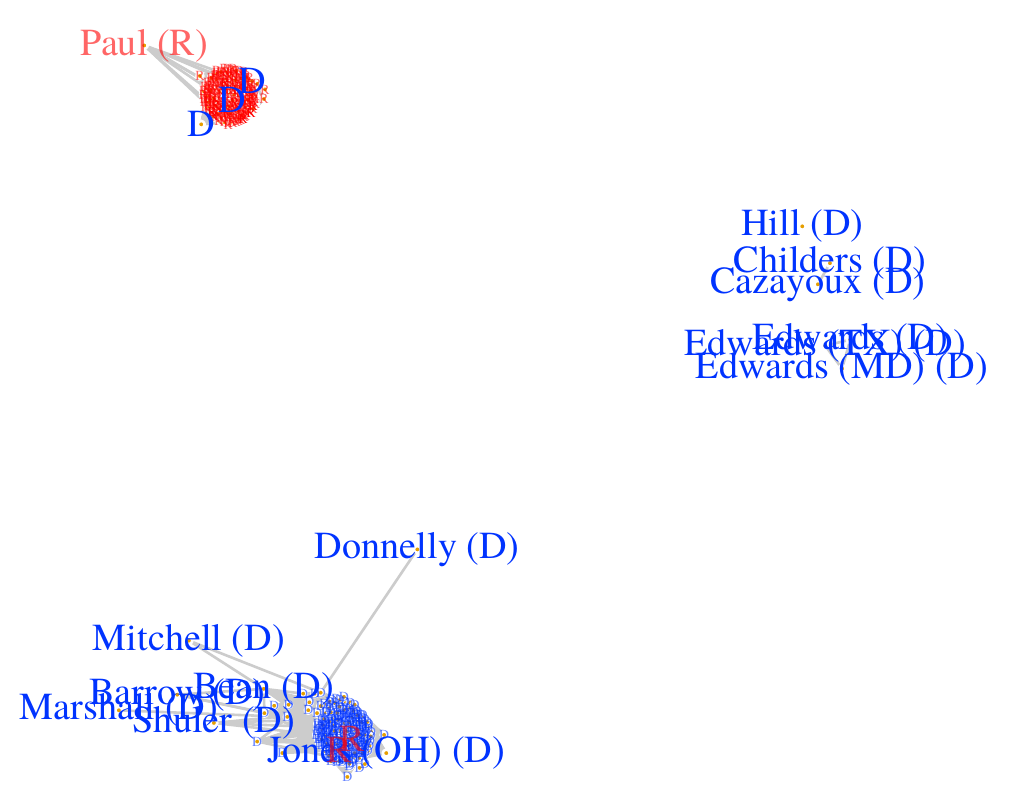}
\caption{The network of Congress members as obtained from voting data for 2008 and the betaMix method. Blue dots represent democratic members, and red dots represent Republicans. Party initial of members who were clustered with the opposite side appear in larger font, and members who were not assigned to a cluster appear with their last name. The other 414 members appear as small dots.}
\label{CongressNetwork2008}
\end{center}
\end{figure}

Based on the degree and clustering coefficient of the nodes, we find two main clusters -- one with 219 Democrats and two Republicans, and one with three Democrats and 195 Republicans. Fourteen members are not assigned to the main two clusters, but some are linked to one of the clusters via at least one edge (e.g. Paul, Mitchell). Figure \ref{CongressNetwork2008} shows that the two clusters are very tight, meaning that most congress members do not deviate much from party lines. Several Democratic members are connected to the blue cluster, but a bit farther from most others in the cluster (e.g. Donnely, Bean, Barrow, etc.) There are six democratic members who appear to have voted sufficiently many times against their party line (or in some case, did not vote enough times to be correlated with their party) and as a result are placed by betaMix outside the blue cluster. Although five members ended up in the cluster of the opposite side, overall the polarization in Congress can hardly be any clearer. It should be noted that 2008 is not the most polarized year. The last decade has been even more polarized according to our method. In fact, in several years in the last decade, the null component is seen to be shifted to the left, and there appear to be fewer correlations which could be explained by chance alone. In such cases, the beta mixture model does not fit the data well, since it assumes that the null set, however small, have a $Beta((\nu-1)/2, 1/2)$ distribution. In the earlier data (1993-1999) the model fits the data very well, and we often find more clusters, and the separation between the two main ones is not so stark. In most cases there are edges between the clusters, indicating that on enough issues the votes were not strictly by party lines (recall that we removed unanimous votes, such as opening the Congress session.) 

\end{document}